\documentclass{article}

\usepackage{PRIMEarxiv}

\usepackage[utf8]{inputenc} 
\usepackage[T1]{fontenc}    
\usepackage[hypertexnames=false]{hyperref}  
\usepackage{url}            
\usepackage{booktabs}       
\usepackage{amsfonts}       
\usepackage{nicefrac}       
\usepackage{microtype}      
\usepackage{fancyhdr}       
\usepackage{graphicx}       
\graphicspath{{figures/}{media/}{pics/}}  

\usepackage{amsmath,amssymb,amsthm,mathtools,bm}
\usepackage{float}
\usepackage{xcolor}
\usepackage{enumitem}

\theoremstyle{plain}
\newtheorem{theorem}{Theorem}[section]
\newtheorem{lemma}[theorem]{Lemma}
\newtheorem{proposition}[theorem]{Proposition}
\newtheorem{corollary}[theorem]{Corollary}

\theoremstyle{definition}
\newtheorem{definition}[theorem]{Definition}

\newtheorem{remark}[theorem]{Remark}

\newcommand{\id}{\mathbb{I}}
\newcommand{\Tr}{\operatorname{Tr}}
\newcommand{\dd}{\mathrm{d}}

\newcommand{\ket}[1]{\lvert #1\rangle}
\newcommand{\bra}[1]{\langle #1\rvert}
\newcommand{\ketbra}[2]{\lvert #1\rangle\!\langle #2\rvert}

\newcommand{\expect}[1]{\left\langle #1\right\rangle}
\newcommand{\norm}[1]{\left\lVert #1\right\rVert}

\title{A Lindblad--Pauli Framework for Coarse-Grained Chaotic Binary-State Dynamics
\thanks{\textit{\underline{Citation}}: Y. Qiu and Q. Zheng, \textbf{A Lindblad--Pauli Framework for Coarse-Grained Chaotic Binary-State Dynamics}. arXiv preprint (2025).}}

\author{
  Yicong Qiu \and Qiye Zheng\thanks{Corresponding author: qiyezheng@ust.hk} \\
  \small Department of Mechanical and Aerospace Engineering, The Hong Kong University of Science and Technology, Hong Kong
}

\date{December 2025}

\begin{document}
\maketitle

\begin{abstract}
Coarse-graining a chaotic bistable oscillator into a binary symbol sequence is a standard reduction, but it often obscures the geometry of the reduced state space and structural constraints of physically meaningful stochastic evolution.
We develop a two-state framework that embeds coarse-grained left/right statistics of the driven Duffing oscillator into a $2\times2$ density-matrix representation and models inter-well switching by a two-rate Gorini--Kossakowski--Sudarshan--Lindblad (GKSL) generator.
For diagonal states the GKSL dynamics reduces to the classical two-state master equation.
The density-matrix language permits an operational ``Bloch half-disk'' embedding with overlap parameter $c(\varepsilon)$ quantifying partition fuzziness; the GKSL model is fitted to diagonal marginals treating $c(\varepsilon)$ as diagnostic.
We derive closed-form solutions, an explicit Kraus representation (generalized amplitude damping with dephasing and rotation), and practical diagnostics for the time-homogeneous first-order Markov assumption (order tests, Chapman--Kolmogorov consistency, run-length statistics, stationarity checks).
When higher-order memory appears, we extend the framework via augmented Markov models, constructing CPTP maps through discrete-time Kraus representations; continuous-time GKSL generators may not exist for all empirical transition matrices.
We provide a numerical pipeline for validating the framework on Duffing simulations.
The density-matrix formalism is an organizational convenience rather than claiming quantum-classical equivalence.
\end{abstract}

\vspace{0.5em}

\section{Introduction}\label{sec:intro}

Chaotic bistable oscillators—systems with two stable attractors separated by a basin boundary and driven by periodic forcing or noise—arise in diverse applications including Josephson junctions \cite{moon1987}, buckled structures, genetic regulatory networks, and neuronal firing models.
When such systems exhibit inter-well transitions on time scales separated from the fast intra-well dynamics, a natural dimensional reduction is to coarse-grain the continuous trajectory into a discrete binary symbol sequence: ``left'' vs.\ ``right'' well occupancy.
This symbolic encoding has proven effective in analyzing chaos \cite{guckenheimer1983}, stochastic resonance \cite{mcnamara1989,gammaitoni1998}, and rare-event transitions in complex systems.

However, standard symbolic dynamics and empirical two-state rate equations face three persistent challenges.
First, \textbf{coarse-graining ambiguity} is typically left implicit: the choice of partition boundary and its effect on inferred transition rates are rarely quantified systematically.
Most implementations adopt a hard partition (e.g., $\mathrm{sign}(x)$) and assume that boundary effects are negligible, without providing a measurable diagnostic for when this assumption holds.
Second, \textbf{Markov closure is often assumed without validation}: discrete-time or continuous-time Markov models are fitted directly to symbol sequences, yet the first-order Markov property (memorylessness) is rarely tested explicitly; higher-order memory effects, time-inhomogeneity, or hidden variables can invalidate these models, leading to spurious rate estimates and unreliable predictions.
Third, \textbf{model extensions lack systematic guarantees}: when the Markov assumption fails (e.g., due to memory of the previous two states), ad hoc extensions are constructed without ensuring that the resulting dynamics remain completely positive and trace-preserving (CPTP)—requirements that are trivial for classical probability vectors but subtle when off-diagonal correlations or soft-partition overlaps are present.

In this work, we develop a unified framework that addresses these challenges by embedding coarse-grained binary statistics into the mathematical structure of quantum two-level systems, while emphasizing that this is a \emph{representational tool} rather than a claim of quantum behavior in classical dynamics.
The Gorini--Kossakowski--Sudarshan--Lindblad (GKSL) theorem \cite{lindblad1976,gorini1976,breuer2002} provides a complete characterization of Markovian CPTP generators in finite dimension; classical two-state master equations are recovered exactly as the diagonal restriction of GKSL dynamics.
The density-matrix language organizes coarse-grained statistics into a $2\times2$ Hermitian matrix with unit trace, where diagonal entries encode occupation probabilities and the off-diagonal entry quantifies partition overlap or measurement uncertainty.
This embedding is not new in principle—classical probability vectors can always be embedded into diagonal density matrices—but the systematic operational implementation for chaotic bistable systems, including explicit soft partitions, falsifiable diagnostics, and CPTP-preserving extensions, has not been developed.

Our approach distinguishes between two density matrices: $\rho_{\mathrm{emb}}(t)$, the \emph{embedded state} constructed directly from phase-space data via soft-partition membership weights $w_L(z)$, $w_R(z)$ (with tunable fuzziness parameter $\varepsilon$), and $\rho_{\mathrm{mod}}(t)$, the \emph{model state} that evolves under a two-rate GKSL generator fitted to the diagonal marginals (populations $p_L$, $p_R$) extracted from $\rho_{\mathrm{emb}}$.
The overlap parameter $c(\varepsilon) = \langle\sqrt{w_L w_R}\rangle$ quantifies boundary fuzziness but is \emph{not} a dynamical variable of the reduced model; it serves as a coarse-graining diagnostic.
The GKSL model governs only population evolution from symbol observations; off-diagonal coherences in $\rho_{\mathrm{mod}}$ represent initial-state uncertainty or mixed preparations and decay to zero at steady state, consistent with the fact that symbols provide no boundary-proximity information.

The primary motivation for adopting GKSL formalism is \textbf{mathematical consistency and automatic CPTP guarantees}: any Markovian dynamics on a two-state system that preserves normalization and nonnegativity of probabilities can be represented in GKSL form (or its discrete-time Kraus counterpart), ensuring that model extensions remain physically consistent by construction.
This is particularly valuable when diagnostics reject the first-order Markov assumption and state-space augmentation is required (e.g., second-order Markov effects necessitate a four-state model on symbol pairs).
Classical rate-equation extensions require manual verification of positivity and normalization constraints at each step; GKSL/Kraus representations inherit these guarantees from the operator-algebraic structure.
We emphasize that this is an \emph{organizational and algebraic} convenience, not a computational speedup: for classical measurement records, classical matrix exponentiation remains the efficient implementation route.

\subsection{Related work and positioning}

Two-state reductions of bistable systems have a long history in nonlinear dynamics and statistical physics.
In the stochastic resonance literature \cite{mcnamara1989,gammaitoni1998}, continuous double-well systems are approximated by two-state master equations with effective switching rates derived from barrier heights, forcing amplitude, and noise intensity via Kramers theory or numerical fitting.
For deterministic chaotic bistability (e.g., the driven Duffing oscillator), a two-state description is an empirical reduction whose adequacy depends on sampling time scale and memory effects.
Our framework differs in three respects: (i) we provide an explicit soft-partition map with quantifiable overlap $c(\varepsilon)$, making coarse-graining resolution testable; (ii) we supply falsifiable diagnostics (order tests, Chapman--Kolmogorov consistency, run-length analysis) with explicit rejection criteria; (iii) we derive closed-form analytical solutions and an equivalence to generalized amplitude damping Kraus operators, connecting the model to quantum-information-theoretic representations.

On the quantum side, the GKSL theorem is foundational for open quantum system dynamics \cite{lindblad1976,gorini1976,breuer2002}.
Recent extensions include universal Lindblad equations \cite{nathan2020universal}, dissipative quantum chaos \cite{kawabata2023symmetry,lange2021random}, many-body systems via Lieb-Robinson bounds \cite{trushechkin2024quantum}, and classical-quantum correspondence \cite{galkowski2025classical,zurek2003decoherence,hernandez2025classical}.
Quantum stochastic walks \cite{whitfield2010} use GKSL-type interpolation between classical and coherent dynamics on graphs.
Quantum simulation of chaotic systems \cite{anand2024simulating,berthusen2020optimal} explores quantum trajectories and superpositions at the microscopic level.
Our framework differs fundamentally in scope: we focus on \emph{classical observables} (left/right occupancy, switching rates, run-length statistics) from deterministic or noise-driven trajectories, using density matrices as an organizational structure for coarse-grained statistics.
The connection between density-matrix methods and phase-space dynamics has been explored via Koopman operator theory \cite{brunton2022modern,li2017extended,mauroy2019koopman} and generalized probabilistic theories \cite{barrett2007,janotta2014}, which provide the mathematical foundation for convex-geometric state-space representations without requiring quantum superposition at the physical level.

\subsection{Contributions and paper organization}

While the formal equivalence between diagonal GKSL dynamics and classical two-state Markov processes is well established, this work provides three operational contributions.
First, we present a \textbf{testable geometric embedding with quantified coarse-graining ambiguity}.
We derive a soft-partition map (Definition~3.1) that constructs $\rho_{\mathrm{emb}}$ from phase-space data with explicit overlap parameter $c(\varepsilon)$, where $\varepsilon$ controls boundary width.
The embedding satisfies three independent consistency criteria: (a)~information-geometric isometry for diagonal states \cite{brody2001,amari2016}, (b)~automatic positivity via Cauchy--Schwarz, (c)~operational interpretation as the Bhattacharyya coefficient.
The resulting Bloch half-disk geometry (Theorem~3.2) imposes the constraint $c^2 \le p_L p_R$, making partition ambiguity explicit and measurable.
This allows systematic selection of $\varepsilon$ by minimizing $c$, matching empirical purity, or requiring $c^2 \ll p_L p_R$ (near-classical limit).

Second, we develop a \textbf{falsifiable diagnostic pipeline with explicit rejection criteria}.
We present four complementary tests (Section~4): (a)~order tests comparing first- vs.\ second-order Markov log-likelihoods via G-test with $\chi^2$ asymptotics; (b)~Chapman--Kolmogorov consistency $\|\widehat{\mathbf{P}}^{(2)} - \widehat{\mathbf{P}}^2\|_F$ with bootstrap confidence intervals; (c)~run-length histograms testing memorylessness via geometric-distribution fits; (d)~windowed stationarity with confidence-interval overlap and heterogeneity $\chi^2$ tests.
Each test makes falsifiable predictions (e.g., $\widehat{\mathbf{P}}^{(2)} \approx \widehat{\mathbf{P}}^2$ for Markov closure, geometric run-length distributions for memorylessness).
Practical guidance for sparse counts (permutation tests, pseudocount smoothing) and stationarity thresholds (Wald intervals, change-point detection) ensures robust application to finite datasets.

Third, we provide \textbf{analytical results and CPTP-preserving extensions}.
We derive closed-form evolution equations (Section~3.1), exact solutions for populations and coherences (Section~3.2), discrete-time Poincar\'e-map formulations with rate-inversion formulas (Section~3.3), and an explicit Kraus representation proving equivalence to generalized amplitude damping composed with dephasing and phase rotation (Section~3.4).
When diagnostics reject first-order Markov closure, we provide systematic state-space augmentation protocols (Section~4.5) that preserve CPTP guarantees by construction; we also give practical criteria for assessing whether continuous-time GKSL generators exist (eigenvalue positivity, matrix-logarithm rate extraction, semigroup consistency) or whether discrete-time Kraus operators should be used instead.

The paper is organized as follows.
Section~2 describes the driven Duffing oscillator, Poincar\'e sampling, binary symbol generation, soft-partition membership functions, and the two-rate GKSL model (our methodological setup).
Section~3 presents analytical results: evolution equations, closed-form solutions, steady-state characterization, discrete-time formulation, and Kraus representation with equivalence proof.
Section~4 develops the diagnostic framework: order tests, Chapman--Kolmogorov consistency, run-length analysis, windowed stationarity, and model-extension protocols.
Section~5 outlines numerical methods and computational validation procedures; detailed numerical examples are provided in Appendices.
Section~6 discusses scope, limitations, interpretive caveats (the two-matrix distinction $\rho_{\mathrm{emb}}$ vs.\ $\rho_{\mathrm{mod}}$, embedding problem for higher-order extensions, relationship to Koopman methods), and future directions.
Section~7 concludes.
Appendices contain derivation details, the complete Kraus equivalence proof, discussion of why pure Hamiltonian evolution cannot model switching, symmetry considerations, a summary of key formulas, and a comprehensive notation reference (Appendix~\ref{app:notation}).

\section{Methods: Model setup}\label{sec:methods}

\subsection{Driven Duffing oscillator}

We consider the driven, damped Duffing oscillator
\begin{equation}\label{eq:duffing}
\ddot{x} + \delta\dot{x} - \alpha x + \beta x^3 = \gamma_0\cos(\omega t),
\end{equation}
where $x$ is displacement, $\delta>0$ is damping, $\alpha>0$ and $\beta>0$ define a symmetric double-well potential, $\gamma_0$ is forcing amplitude, and $\omega$ is forcing frequency.
The potential energy is
\begin{equation}\label{eq:potential}
V(x) = -\frac{1}{2}\alpha x^2 + \frac{1}{4}\beta x^4,
\end{equation}
with stable equilibria at $x_\pm=\pm\sqrt{\alpha/\beta}$ and barrier height $\Delta V=\alpha^2/(4\beta)$.

\begin{remark}[Optional stochastic extension]
If one adds Gaussian white noise $\sqrt{2D}\,\xi(t)$ to the forcing, the switching statistics often become closer to Poisson at suitable sampling scales, which can strengthen the empirical case for a time-homogeneous Markov reduction:
\begin{equation}\label{eq:duffing_noise}
\ddot{x} + \delta\dot{x} - \alpha x + \beta x^3 = \gamma_0\cos(\omega t) + \sqrt{2D}\,\xi(t).
\end{equation}
The deterministic ($D=0$) development below still applies; the noise term simply changes the effective rates and can broaden the boundary region, which can be modeled by the soft-partition parameter $\varepsilon$ (Section~\ref{sec:coarse_graining}).
\end{remark}

\subsection{Poincar\'e sampling and symbol sequence}

Let $T=2\pi/\omega$ be the forcing period, and define sampling times $t_n=nT$.
The simplest binary coding uses the sign of $x(t_n)$:
\begin{definition}[Binary symbol sequence]\label{def:symbol}
Define $S_n\in\{L,R\}$ by
\begin{equation}
S_n =
\begin{cases}
L, & x(t_n)<0,\\
R, & x(t_n)>0.
\end{cases}
\end{equation}
\end{definition}
This yields a discrete-time binary process $\{S_n\}_{n\ge 0}$ (after discarding transients).

\begin{figure}[t]
\centering
\includegraphics[width=0.99\textwidth]{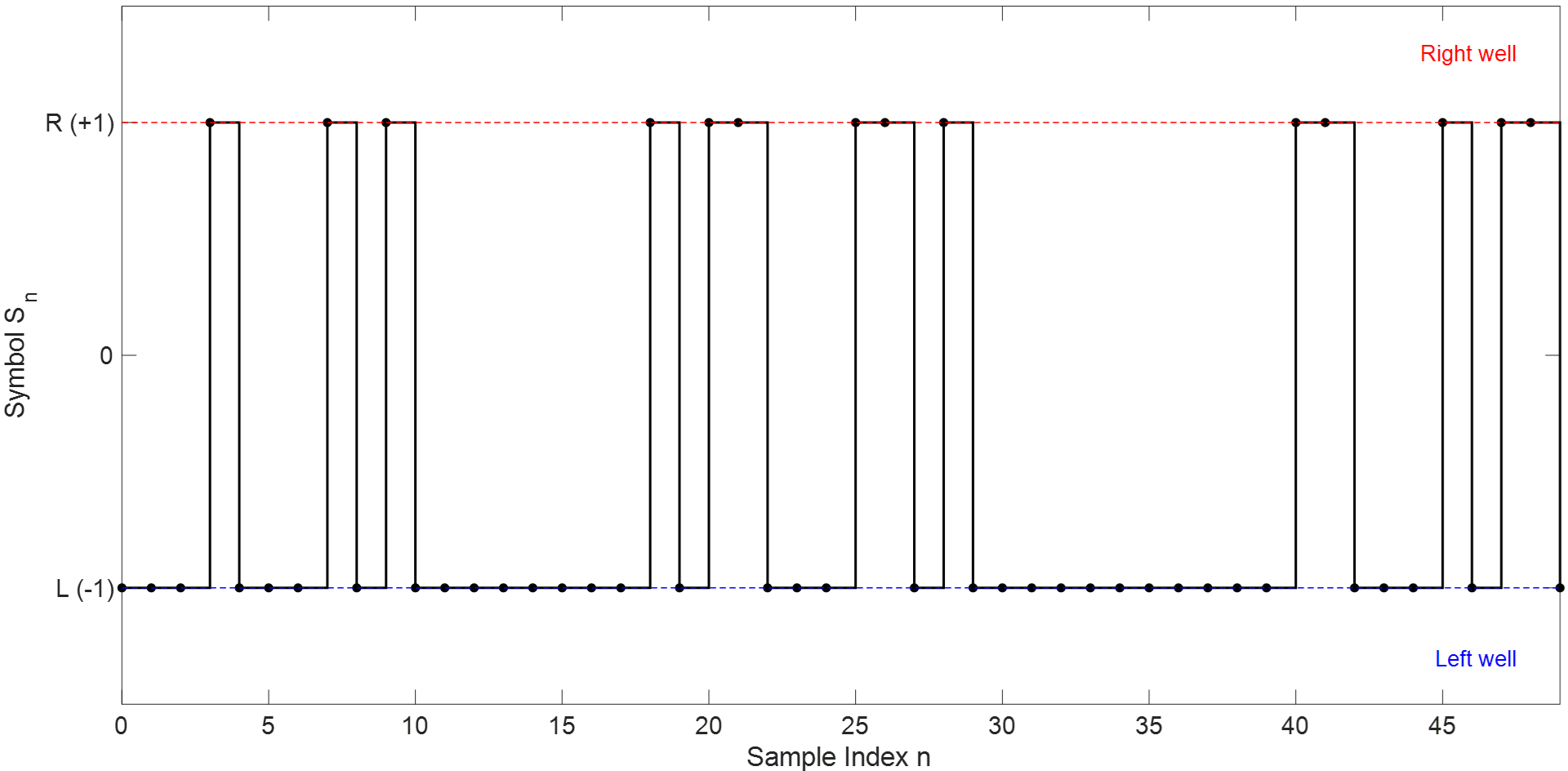}
\caption{Example segment of the binary symbol sequence $\{S_n\}$ (or equivalently the sign of $x(t_n)$) illustrating dwell times and switching events.}
\label{fig:symbol_sequence}
\end{figure}

In practice, one may use more refined boundary functions than $x\mapsto \mathrm{sign}(x)$ (for instance, an approximate separatrix on the stroboscopic section), and one may also consider windowed statistics (e.g.\ average over $t\in[t_n,t_n+\Delta]$) rather than pointwise sampling.
Figure~\ref{fig:symbol_sequence} illustrates an example segment of the resulting binary symbol sequence.

\subsection{Soft-partition embedding and Bloch geometry}\label{sec:coarse_graining}

To quantify the effect of coarse-graining resolution on the reduced statistics, we now formalize the embedding of phase-space data into a $2\times2$ density matrix with an explicit overlap parameter $c(\varepsilon)$ that characterizes boundary fuzziness.

\subsubsection{Two-state basis and Pauli operators}

We introduce the computational basis
\begin{equation}
\ket{L}=\begin{pmatrix}1\\0\end{pmatrix},\qquad
\ket{R}=\begin{pmatrix}0\\1\end{pmatrix},
\end{equation}
together with the Pauli matrices
\begin{equation}\label{eq:pauli}
\sigma_x=\begin{pmatrix}0&1\\1&0\end{pmatrix},\quad
\sigma_y=\begin{pmatrix}0&-i\\ i&0\end{pmatrix},\quad
\sigma_z=\begin{pmatrix}1&0\\0&-1\end{pmatrix}.
\end{equation}
We also use the raising/lowering operators
\begin{equation}\label{eq:sigma_pm}
\sigma_+ = \ketbra{R}{L} = \begin{pmatrix}0&0\\1&0\end{pmatrix},\qquad
\sigma_- = \ketbra{L}{R} = \begin{pmatrix}0&1\\0&0\end{pmatrix}.
\end{equation}

\subsection{Density matrices and the Bloch representation}

A (qubit) density matrix $\rho$ is a $2\times2$ Hermitian positive semidefinite matrix with unit trace.
Equivalently, any such $\rho$ can be written in Bloch form
\begin{equation}\label{eq:bloch}
\rho=\frac{1}{2}\Bigl(\id+m_x\sigma_x+m_y\sigma_y+m_z\sigma_z\Bigr)
=\frac{1}{2}\begin{pmatrix}
1+m_z & m_x-im_y\\
m_x+im_y & 1-m_z
\end{pmatrix},
\end{equation}
with Bloch vector $\bm{m}=(m_x,m_y,m_z)$ satisfying $\|\bm{m}\|\le 1$.

\begin{remark}[Classical diagonal states]
A purely classical two-state distribution $(p_L,p_R)$ corresponds to a diagonal density matrix
\begin{equation}\label{eq:diagonal}
\rho=\begin{pmatrix}p_L&0\\0&p_R\end{pmatrix}
=\frac{1}{2}(\id+m_z\sigma_z),\qquad m_z=p_L-p_R.
\end{equation}
In this case only $m_z$ is needed.
\end{remark}

\subsection{Soft partitions and an operational overlap parameter}

\paragraph{Why introduce off-diagonal terms?}
A single deterministic Duffing trajectory is never ``in a superposition'' of wells.
However, a \emph{coarse-grained} description may be uncertain when the trajectory is close to the boundary or when observations are aggregated over a finite window.
We represent this uncertainty by soft membership functions and an embedding that produces a real-symmetric $2\times2$ matrix with a nonnegative off-diagonal entry.

\begin{definition}[Soft partition / membership functions]\label{def:soft_partition}
Let $z=(x,\dot{x})$ denote a point on a chosen Poincar\'e section.
Introduce $w_L(z)\in[0,1]$ and $w_R(z)=1-w_L(z)$ with
\begin{equation}\label{eq:soft_wL}
w_L(z)=\frac{1}{2}\Bigl(1-\tanh\!\bigl(\tfrac{g(z)}{\varepsilon}\bigr)\Bigr),\qquad
w_R(z)=\frac{1}{2}\Bigl(1+\tanh\!\bigl(\tfrac{g(z)}{\varepsilon}\bigr)\Bigr),
\end{equation}
where $g(z)$ is a signed boundary function (zero level-set approximates the separatrix on the section) and $\varepsilon>0$ controls fuzziness. The hard partition is recovered as $\varepsilon\to0^+$.
For exposition we often take $g(z)=x$.
\end{definition}

\begin{definition}[Pointwise embedded two-state vector]\label{def:psi_of_z}
Associate to each $z$ the real-amplitude vector
\begin{equation}\label{eq:psi_z}
\ket{\psi(z)}=\sqrt{w_L(z)}\,\ket{L}+\sqrt{w_R(z)}\,\ket{R}.
\end{equation}
This is an embedding device that encodes partial membership induced by coarse-graining; it is not a microscopic quantum state of the Duffing oscillator.
\end{definition}

\begin{remark}[Why the square-root embedding?]\label{rem:sqrt_choice}
The choice of square-root weighting in \eqref{eq:psi_z} is not arbitrary but is motivated by three independent considerations from information geometry, positivity constraints, and statistical overlap measures.

\textbf{(1) Hellinger distance and information geometry.}
The Hellinger distance between probability distributions $p$ and $q$ is $H(p,q)=\sqrt{1-\int\sqrt{p(x)q(x)}\,dx}$.
The square-root embedding connects the Bloch-sphere geometry to the Fisher-Rao information geometry \textit{for the classical (diagonal/commuting) submanifold} \cite{brody2001,amari2016}, establishing a close connection between the Fubini-Study metric and the Fisher-Rao metric for this restricted case.
This ensures that geometric notions (distance, curvature) have consistent interpretations across quantum and classical descriptions.

\textbf{(2) Automatic positivity via Cauchy-Schwarz.}
With the square-root embedding, the off-diagonal element $c(t)=\int\mu_t(z)\sqrt{w_L(z)w_R(z)}\,dz$ satisfies $c^2\le p_L p_R$ by the Cauchy-Schwarz inequality applied to $\int\sqrt{w_L}\sqrt{w_R}$.
This \emph{automatically} ensures that the resulting density matrix is positive semidefinite without additional constraints.
Alternative embeddings (e.g., $c=\int w_L w_R$ or $c=\int w_L^{1/3}w_R^{2/3}$) would require manual verification of positivity at each time step.

\textbf{(3) Bhattacharyya overlap coefficient.}
Define induced sub-densities $p_L(z) = \mu_t(z) w_L(z)$ and $p_R(z) = \mu_t(z) w_R(z)$.
Then
\[
c(t) = \int \sqrt{p_L(z) p_R(z)}\,dz
\]
is precisely the Bhattacharyya coefficient (Hellinger affinity) between $p_L$ and $p_R$ \cite{kailath1967}.
This standard measure of distributional overlap in statistics and pattern recognition gives $c(t)$ a direct operational interpretation: it quantifies the overlap between left-well and right-well membership distributions at resolution $\varepsilon$.

These three principles---information-geometric isometry, automatic positivity, and statistical interpretability---converge to the same embedding, demonstrating that the square-root choice is principled rather than ad hoc.
Alternative embeddings are possible (e.g., direct diagonal: $\rho=\mathrm{diag}(p_L,p_R)$ with $c=0$), but forfeit either geometric naturalness or the ability to represent partition fuzziness.
\end{remark}

\begin{definition}[Ensemble or window-reduced state]\label{def:rho_from_mu}
Let $\mu_t(z)$ be a probability density (e.g.\ an ensemble of initial conditions) or an empirical measure derived from a time window.
Define the embedded state
\begin{equation}\label{eq:rho_from_mu}
\rho_{\mathrm{emb}}(t;\varepsilon,\Delta)=\int \mu_t(z)\,\ket{\psi(z)}\bra{\psi(z)}\,\dd z.
\end{equation}
\end{definition}

\begin{proposition}[Matrix elements and overlap]\label{prop:rho_elements_overlap}
With $\rho_{\mathrm{emb}}(t;\varepsilon,\Delta)$ as in \eqref{eq:rho_from_mu},
\begin{align}
[\rho_{\mathrm{emb}}]_{LL}(t) &= \int \mu_t(z)\,w_L(z)\,\dd z=:p_L(t),\\
[\rho_{\mathrm{emb}}]_{RR}(t) &= \int \mu_t(z)\,w_R(z)\,\dd z=:p_R(t)=1-p_L(t),\\
[\rho_{\mathrm{emb}}]_{LR}(t) &= \int \mu_t(z)\,\sqrt{w_L(z)w_R(z)}\,\dd z=:c(t)\in[0,\tfrac12].
\end{align}
The upper bound $c(t) \le \frac{1}{2}$ follows from the constraint $w_L + w_R = 1$ and the AM-GM inequality: $\sqrt{w_L w_R} \le \frac{w_L + w_R}{2} = \frac{1}{2}$, with equality only when $w_L = w_R = \frac{1}{2}$ (maximum boundary fuzziness).
Thus
\begin{equation}\label{eq:rho_real_form}
\rho_{\mathrm{emb}}(t)=\begin{pmatrix}p_L(t)&c(t)\\ c(t)&1-p_L(t)\end{pmatrix}.
\end{equation}
Moreover $c(t)\to0$ in the hard-partition limit $\varepsilon\to0^+$.
\end{proposition}

\begin{definition}[Windowed empirical measure]\label{def:window_measure}
Given a trajectory $z(t)$ on the section and a window length $\Delta>0$, define
\begin{equation}
\mu_{t,\Delta}(A)=\frac{1}{\Delta}\int_{t}^{t+\Delta}\mathbf{1}_{\{z(s)\in A\}}\,\dd s
\end{equation}
for measurable sets $A$.
\end{definition}

\begin{remark}[Discrete Poincar\'e sampling]
For discrete Poincar\'e samples $z_n = z(t_n)$ with $t_n = nT$, the windowed measure becomes
\[
\mu_{n,M}(A) = \frac{1}{M}\sum_{j=0}^{M-1} \mathbf{1}_{\{z_{n+j}\in A\}},
\]
where $M$ is the window length in sampling periods.
\end{remark}

\begin{remark}[Operational meaning of $c(t)$]
Under the windowed measure $\mu_{t,\Delta}$, $c(t;\Delta)$ increases when the window overlaps the inter-well boundary region where $w_Lw_R>0$.
Thus $c(t)$ quantifies \emph{how much of the coarse-grained mass is near the decision boundary}.
In empirical work, one may estimate $c$ from sampled points by
\[
\hat c=\frac{1}{N}\sum_{n=1}^N \sqrt{w_L(z(t_n))w_R(z(t_n))},
\]
with $w_L,w_R$ fixed by the chosen coarse-graining and resolution parameter $\varepsilon$.
\end{remark}

\begin{remark}[c(t) is not a dynamical variable]
The overlap parameter $c(t)$ characterizes the coarse-graining resolution but is \emph{not} a dynamical observable of the reduced model.
The GKSL model (Section~\ref{sec:gksl_model}) governs the evolution of populations $p_L(t), p_R(t)$ from symbol observations; the off-diagonal entries $\rho_{LR}$ in the density-matrix representation encode initial-state uncertainty or measurement resolution, but cannot be inferred from symbol data alone.
In practice, $c(\varepsilon)$ serves as a diagnostic for partition quality: smaller $c$ indicates sharper boundaries and better approximation of the classical limit (diagonal states).
The GKSL steady state $\rho^\infty$ is diagonal (Theorem~\ref{thm:steady}), which is consistent because symbols provide no boundary-proximity information.
\end{remark}

\paragraph{Two density matrices: embedding vs.\ dynamical model.}
\label{para:two_rhos}
We distinguish two density-matrix objects throughout this work.
The first is the \emph{embedded state} $\rho_{\mathrm{emb}}(t;\varepsilon,\Delta)$ constructed from phase-space data via the soft-partition map \eqref{eq:psi_z} and \eqref{eq:rho_from_mu}.
This state is a functional of the phase-space distribution $\mu_t(z)$ and the partition parameters $(\varepsilon,\Delta)$; its off-diagonal element $[\rho_{\mathrm{emb}}]_{LR} = c(t;\varepsilon)$ quantifies boundary overlap and serves as a coarse-graining diagnostic.
The second is the \emph{model state} $\rho_{\mathrm{mod}}(t)$ governed by the GKSL equation (Section~\ref{sec:gksl_model}).
This state represents the reduced two-state dynamics inferred from symbol observations $\{S_n\}$.
In practice, the GKSL model is fitted and validated against the \emph{diagonal marginals} of $\rho_{\mathrm{emb}}$, namely the populations $p_L(t), p_R(t)$.
The off-diagonal terms of $\rho_{\mathrm{mod}}$ represent initial-state uncertainty or mixed preparations; they are not directly measurable from symbol data and decay to zero at steady state (Theorem~\ref{thm:steady}).

The overlap parameter $c(t;\varepsilon)$ characterizes the quality of the coarse-graining (how much boundary region is ambiguous) but is \emph{not} a closed dynamical variable of $\rho_{\mathrm{mod}}$.
The two-state GKSL framework models population evolution from symbols; boundary-proximity information in $c(\varepsilon)$ requires phase-space or soft-membership data beyond the symbol sequence.

\subsection{Bloch half-disk geometry}

\begin{theorem}[Bloch half-disk state space]\label{thm:bloch_disk}
A real-symmetric matrix of the form \eqref{eq:rho_real_form} (i.e., $\rho_{\mathrm{emb}}$ with off-diagonal $c(t)$) is positive semidefinite if and only if
\begin{equation}\label{eq:disk_condition}
c(t)^2\le p_L(t)\bigl(1-p_L(t)\bigr).
\end{equation}
Equivalently, writing $\rho_{\mathrm{emb}}=\frac{1}{2}(\id+m_x\sigma_x+m_z\sigma_z)$ with $m_y=0$, one has $m_x=2c\ge0$ and
\begin{equation}\label{eq:bloch_disk_set}
\mathcal S_{\mathrm{half\text{-}disk}}
=\Bigl\{\rho:\ m_y=0,\ m_x\ge 0,\ m_x^2+m_z^2\le 1\Bigr\}.
\end{equation}
The classical two-state simplex corresponds to the diameter $m_x=0$ (diagonal states).
\end{theorem}

\begin{proof}
For $\rho=\begin{pmatrix}p&c\\c&1-p\end{pmatrix}$, $\Tr(\rho)=1$ and $\det(\rho)=p(1-p)-c^2$.
Positive semidefiniteness is equivalent to $\det(\rho)\ge0$, yielding \eqref{eq:disk_condition}.
Using $m_z=2p-1$ and $m_x=2c$ gives $m_x^2+m_z^2\le1$ with $m_y=0$.
Nonnegativity of $c$ follows from the integral definition in Proposition~\ref{prop:rho_elements_overlap}.
\end{proof}

\begin{remark}[State-space geometry vs accessible measurements]
The half-disk describes the \emph{embedded state space}.
If the only accessible readout is $L/R$ occupancy (a $\sigma_z$-type measurement), then states with the same $p_L$ are operationally indistinguishable.
To operationalize $c(t)$ one must define additional observables tied to the boundary region (e.g.\ short-time boundary-crossing indicators or augmented coarse-grainings).
\end{remark}

\begin{remark}[Closure of the half-disk]
The half-disk state space is closed under evolution only when $\Omega=0$ (no energy splitting) and initial states are real-symmetric.
For $\Omega\neq 0$, the Hamiltonian term induces phase rotation, causing $\rho_{LR}(t)$ to acquire an imaginary component and $m_y\neq 0$, thereby leaving the real half-disk.
In symmetric bistable systems (e.g., Duffing potential with $V(-x)=V(x)$, no bias terms), physical symmetry implies $\Delta E=0$ and hence $\Omega=0$.
\end{remark}

\subsection{Two-rate GKSL generator}\label{sec:gksl_model}

Having embedded coarse-grained statistics into a density-matrix representation, we now construct a dynamical model for inter-well switching.
The model acts on the two-state Hilbert space and governs the evolution of $\rho_{\mathrm{mod}}(t)$, which we fit to the diagonal marginals (populations) extracted from $\rho_{\mathrm{emb}}$.

\subsubsection{Why GKSL?}

A purely Hamiltonian (unitary) evolution cannot change the populations of a diagonal state when the Hamiltonian is diagonal in the $\{\ket{L},\ket{R}\}$ basis.
Since inter-well switching \emph{is} a population-transfer phenomenon, an effective reduced model must include dissipative (jump) terms.
Appendix~\ref{app:hamiltonian_fails} gives a formal proof and also explains why common non-Hermitian ``dissipative Hamiltonian'' tricks lead to nonlinear (non-CPTP) dynamics.

\subsubsection{GKSL form and model specification}

\begin{definition}[GKSL master equation]\label{def:gksl}
A Markovian, trace-preserving, completely positive master equation on density matrices takes the GKSL form
\begin{equation}\label{eq:lindblad_general}
\frac{\dd\rho}{\dd t}
=-\frac{i}{\hbar}[H,\rho]
+\sum_k\left(L_k\rho L_k^\dagger-\frac12\{L_k^\dagger L_k,\rho\}\right),
\end{equation}
where $H=H^\dagger$ and $\{A,B\}=AB+BA$.
\end{definition}

\begin{definition}[Two-state Lindblad model]\label{def:model}
We use
\begin{align}
H &= -\frac{\Delta E}{2}\sigma_z,\label{eq:H_model}\\
L_+ &= \sqrt{k_{LR}}\,\sigma_+,\label{eq:L_plus}\\
L_- &= \sqrt{k_{RL}}\,\sigma_-,\label{eq:L_minus}\\
L_d &= \sqrt{\kappa}\,\sigma_z\quad\text{(optional pure dephasing)},\label{eq:L_d}
\end{align}
with rates $k_{LR},k_{RL}\ge0$, dephasing rate $\kappa\ge0$, and $\Omega=\Delta E/\hbar$.
\end{definition}

\begin{remark}[Interpretation]
$L_+$ and $L_-$ implement population transfers $L\to R$ and $R\to L$ at rates $k_{LR}$ and $k_{RL}$.
The $L_d$ term damps off-diagonal elements without changing populations.
The GKSL model acts on the reduced two-state space; off-diagonal terms $\rho_{LR}$ in this model represent initial-state uncertainty or mixed preparations, not the coarse-graining overlap $c(\varepsilon)$ defined in Section~\ref{sec:coarse_graining}.
Because symbol observations contain no boundary-proximity information, $\rho_{LR}$ cannot be measured from symbol data and decays to zero at steady state (Section~\ref{sec:solutions}).
The overlap parameter $c(\varepsilon)$ is instead a \emph{coarse-graining diagnostic} that quantifies partition fuzziness for a given $\varepsilon$ (Figure~\ref{fig:c_vs_epsilon}).
Throughout, we use $\rho_{\mathrm{emb}}(t;\varepsilon)$ to denote the embedded state from phase-space data (whose off-diagonal contains $c(\varepsilon)$, see \S\ref{para:two_rhos}) and $\rho_{\mathrm{mod}}(t)$ to denote the state governed by this GKSL equation.
The model predicts the diagonal part: $[\rho_{\mathrm{mod}}]_{LL} \approx p_L(t) = [\rho_{\mathrm{emb}}]_{LL}$.
\end{remark}

\begin{remark}[Default: symmetric case with $\Omega=0$]
For symmetric bistable potentials (e.g., Duffing with no bias or asymmetry terms), physical symmetry $V(-x)=V(x)$ implies $\Delta E=0$, yielding $\Omega=0$.
In this case, the half-disk representation remains closed and $m_y=0$ throughout evolution.
Nonzero $\Omega$ corresponds to biased or asymmetric potentials and induces phase rotation that mixes real and imaginary components of $\rho_{LR}$.
\end{remark}

\subsection{Classical limit on diagonal states}

\begin{proposition}[Reduction to the classical two-state master equation]\label{prop:classical_limit}
If $\rho_{\mathrm{mod}}(t)$ is diagonal in the $\{\ket{L},\ket{R}\}$ basis, i.e.\ $\rho_{\mathrm{mod}}=\mathrm{diag}(p_L,1-p_L)$, then the GKSL model of Definition~\ref{def:model} yields
\begin{equation}\label{eq:classical_master}
\frac{\dd p_L}{\dd t}=-k_{LR}p_L+k_{RL}(1-p_L),
\end{equation}
which is the standard continuous-time two-state (telegraph) master equation with time-homogeneous rates.
\end{proposition}

\begin{proof}
With $\rho_{LR}=0$, the Hamiltonian commutator vanishes and the dephasing dissipator vanishes on diagonal entries.
The jump dissipators generated by $L_\pm$ produce the gain/loss terms shown in \eqref{eq:classical_master}; see Theorem~\ref{thm:evolution} below or Appendix~\ref{app:evolution_derivation}.
\end{proof}

\section{Results: Analytical characterization}\label{sec:results_analytical}

Having specified the GKSL model, we now derive explicit evolution equations, closed-form solutions, discrete-time formulations, and an equivalence to Kraus operators.
These analytical results provide the mathematical foundation for model fitting and diagnostic testing.

\subsection{Evolution equations and Bloch dynamics}\label{sec:evolution_equations}

For the model state $\rho_{\mathrm{mod}}(t)$, write
\[
\rho_{\mathrm{mod}}=\begin{pmatrix}\rho_{LL}&\rho_{LR}\\\rho_{RL}&\rho_{RR}\end{pmatrix},\qquad
\rho_{RL}=\rho_{LR}^*,\qquad \rho_{LL}+\rho_{RR}=1.
\]

\begin{theorem}[Explicit evolution equations]\label{thm:evolution}
For the model in Definition~\ref{def:model},
\begin{align}
\frac{\dd\rho_{LL}}{\dd t} &= -k_{LR}\rho_{LL}+k_{RL}\rho_{RR},\label{eq:rhoLL}\\
\frac{\dd\rho_{RR}}{\dd t} &= +k_{LR}\rho_{LL}-k_{RL}\rho_{RR},\label{eq:rhoRR}\\
\frac{\dd\rho_{LR}}{\dd t} &= \left(i\Omega-\frac{\Gamma}{2}-2\kappa\right)\rho_{LR},\label{eq:rhoLR}
\end{align}
where $\Gamma=k_{LR}+k_{RL}$.
\end{theorem}

\begin{remark}
A detailed step-by-step derivation is provided in Appendix~\ref{app:evolution_derivation}. The result is a linear ODE system: the populations follow a closed two-state rate equation and the off-diagonal term undergoes damped rotation with decay rate $\Gamma/2+2\kappa$.
\end{remark}

\subsubsection{Bloch-vector form}

Using $\rho_{\mathrm{mod}}=\frac12(\id+\bm{m}\cdot\bm{\sigma})$ with $\bm{m}=(m_x,m_y,m_z)$, one obtains
\begin{corollary}[Bloch-vector evolution]\label{cor:bloch_evolution}
The Bloch components satisfy
\begin{align}
\dot m_x &= \Omega m_y-\left(\frac{\Gamma}{2}+2\kappa\right)m_x,\label{eq:mx}\\
\dot m_y &= -\Omega m_x-\left(\frac{\Gamma}{2}+2\kappa\right)m_y,\label{eq:my}\\
\dot m_z &= -\Gamma(m_z-m_z^\infty),\qquad m_z^\infty=\frac{k_{RL}-k_{LR}}{\Gamma}.\label{eq:mz}
\end{align}
When $\Omega=0$ (symmetric case), the evolution remains in the $(m_x, m_z)$ plane with $m_y=0$, preserving the half-disk embedding from Section~\ref{sec:coarse_graining}.
\end{corollary}

\subsection{Closed-form solutions and steady state}\label{sec:solutions}

\subsubsection{Steady state}

\begin{theorem}[Steady state of the GKSL model]\label{thm:steady}
Assume $\Gamma = k_{LR} + k_{RL} > 0$.
Then the unique stationary state of $\rho_{\mathrm{mod}}$ under the GKSL dynamics \eqref{eq:lindblad_general} is diagonal,
\begin{equation}\label{eq:steady_state}
\rho_{\mathrm{mod}}^\infty=\begin{pmatrix}p_L^\infty&0\\0&p_R^\infty\end{pmatrix},
\qquad
p_L^\infty=\frac{k_{RL}}{\Gamma},\quad p_R^\infty=\frac{k_{LR}}{\Gamma}.
\end{equation}
This is consistent with $[\rho_{\mathrm{emb}}]_{LR} = c(\varepsilon) > 0$ for soft partitions, because $c(\varepsilon)$ characterizes the embedding geometry, not a prediction of the symbol-based model.
\end{theorem}

\subsection{Population relaxation}

\begin{theorem}[Population solution for diagonal initial states]\label{thm:pop_solution}
For $\rho_{\mathrm{mod}}(0)=\mathrm{diag}(p_0,1-p_0)$,
\begin{equation}\label{eq:pop_solution}
p_L(t)=p_L^\infty+(p_0-p_L^\infty)e^{-\Gamma t},
\end{equation}
and $\rho_{\mathrm{mod}}(t)$ remains diagonal.
\end{theorem}

The characteristic relaxation time is $\tau_{\mathrm{rel}}=1/\Gamma$.

\subsection{Off-diagonal decay}

\begin{theorem}[Off-diagonal evolution]\label{thm:coherence}
For general initial $\rho_{LR}(0)$,
\begin{equation}\label{eq:coherence_solution}
\rho_{LR}(t)=\rho_{LR}(0)\exp\!\left[\left(i\Omega-\frac{\Gamma}{2}-2\kappa\right)t\right].
\end{equation}
\end{theorem}

This defines an effective decay time $\tau_{\mathrm{off}}=1/(\Gamma/2+2\kappa)$ for off-diagonal components.
The off-diagonal decay describes how initial preparation uncertainty or mixed states evolve under the reduced GKSL dynamics.
This is \emph{distinct} from the coarse-graining overlap $c(\varepsilon)$, which characterizes the partition geometry at a given resolution $\varepsilon$ (Section~\ref{sec:coarse_graining}).
At steady state, $\rho_{LR}^\infty = 0$ because symbol data provide no boundary information; the nonzero $c(\varepsilon)>0$ for soft partitions reflects the embedding choice, not a dynamical prediction.

\subsection{Discrete-time Poincar\'e map and transition matrix}

When observations are made once per forcing period $T$, one can work with a discrete-time transition matrix.
\begin{theorem}[Discrete-time transition probabilities]\label{thm:discrete}
Sampling at interval $T$, the populations satisfy
\begin{equation}\label{eq:discrete_update}
p_L^{(n+1)}=p_L^\infty+\bigl(p_L^{(n)}-p_L^\infty\bigr)e^{-\Gamma T}.
\end{equation}
Equivalently,
\begin{equation}\label{eq:transition_matrix}
\begin{pmatrix}p_L\\p_R\end{pmatrix}_{n+1}
=
\begin{pmatrix}
1-P_{LR} & P_{RL}\\
P_{LR} & 1-P_{RL}
\end{pmatrix}
\begin{pmatrix}p_L\\p_R\end{pmatrix}_n,
\end{equation}
with
\begin{equation}\label{eq:Pij}
P_{LR}=p_R^\infty(1-e^{-\Gamma T}),\qquad
P_{RL}=p_L^\infty(1-e^{-\Gamma T}).
\end{equation}
\end{theorem}

\subsection{Rate estimation from symbol data}

Given a long symbol sequence $\{S_n\}$, define empirical one-step transition probabilities
\[
\hat P_{LR}=\frac{\#(L\to R)}{\#(\text{steps starting in }L)},\qquad
\hat P_{RL}=\frac{\#(R\to L)}{\#(\text{steps starting in }R)}.
\]
Under the time-homogeneous two-state Markov hypothesis at sampling interval $T$, one may invert \eqref{eq:Pij}:
\begin{equation}\label{eq:Gamma_hat}
\hat\Gamma=-\frac{1}{T}\ln\!\left(1-\hat P_{LR}-\hat P_{RL}\right),
\end{equation}
\begin{equation}\label{eq:k_hat}
\hat k_{LR}=\frac{\hat P_{LR}}{\hat P_{LR}+\hat P_{RL}}\hat\Gamma,\qquad
\hat k_{RL}=\frac{\hat P_{RL}}{\hat P_{LR}+\hat P_{RL}}\hat\Gamma.
\end{equation}
Section~\ref{sec:markov_tests} emphasizes that these formulas should be used only after Markov diagnostics support the assumed closure at the chosen sampling scale.

\section{Results: Diagnostic framework}\label{sec:markov_tests}

The analytical results derived in Section~\ref{sec:results_analytical} rely on the assumption that the coarse-grained symbol sequence $\{S_n\}$ satisfies a time-homogeneous first-order Markov property.
This section develops a complete diagnostic framework with four complementary tests that can falsify this assumption, provides practical guidance for sparse data and threshold selection, and outlines systematic extensions when higher-order memory is detected.

\subsection{Hypotheses and test structure}

\begin{definition}[First-order Markov property]
$\{S_n\}$ is first-order Markov if for all $i,j\in\{L,R\}$,
\[
\Pr(S_{n+1}=j\mid S_n=i,S_{n-1},\dots)=\Pr(S_{n+1}=j\mid S_n=i)=:P_{ij}.
\]
\end{definition}

\begin{definition}[Time homogeneity]
The chain is time homogeneous if $P_{ij}$ does not depend on $n$ after transients.
\end{definition}

\subsection{Order test: does conditioning on $S_{n-1}$ change transitions?}

Define counts
\[
N_{kij}=\#\{n:\ S_{n-1}=k,\ S_n=i,\ S_{n+1}=j\},\quad
N_{ki}=\sum_{j}N_{kij},\quad
N_{ij}=\sum_{k}N_{kij}.
\]
Define empirical probabilities
\[
\widehat P_{ij}=\frac{N_{ij}}{\sum_{j'}N_{ij'}},\qquad
\widehat P_{ij\mid k}=\frac{N_{kij}}{N_{ki}}.
\]
A standard likelihood-ratio statistic for first-order vs second-order structure is
\begin{equation}\label{eq:g_test}
G=2\sum_{k,i,j}N_{kij}\ln\!\left(\frac{\widehat P_{ij\mid k}}{\widehat P_{ij}}\right),
\end{equation}
which is asymptotically $\chi^2$ distributed under the null (large-sample regime).
If $G$ is large relative to the appropriate degrees of freedom, first-order Markovity is rejected.
For the G-test comparing first-order vs.\ second-order Markov models, the degrees of freedom are
\begin{equation}\label{eq:order_test_df}
  \mathrm{df} = d(d-1)^2,
\end{equation}
where $d$ is the number of states. For the binary case ($d=2$), $\mathrm{df}=2$.
At significance level $\alpha=0.05$, the critical value is $\chi^2_{2,0.05}\approx 5.99$.
The empirical $p$-value is computed as $p = \Pr(\chi^2_2 \ge G)$. The null hypothesis
of first-order Markovity is rejected if $p < \alpha$.

\begin{remark}[Implementation and sparse-count handling]
In computing $G$, terms with $N_{kij} = 0$ are omitted (the limit $0\log 0 = 0$).
When some $N_{kij}$ are small (say $<5$), the asymptotic $\chi^2$ approximation may be unreliable.
Three practical remedies are recommended:
\begin{enumerate}[label=(\alph*), leftmargin=2em]
\item \textbf{Pseudocount smoothing:} Add a small Laplace pseudocount $\delta \in [0.5, 1]$ to all counts: $N_{kij} \to N_{kij} + \delta$. Report the $p$-value for $\delta=0$ (no smoothing) and $\delta=0.5$ to assess sensitivity.
\item \textbf{Permutation test:} If many cells have $N_{kij} < 5$, replace the $\chi^2$ reference distribution with a permutation test: shuffle the symbol sequence $\{S_n\}$ randomly (preserving marginals), recompute $G^{\mathrm{perm}}$, and repeat 1000--10000 times to build an empirical null distribution. Reject if the observed $G$ exceeds the 95th percentile of $G^{\mathrm{perm}}$.
\item \textbf{Bootstrap confidence intervals:} Use block bootstrap (preserving temporal dependence) to obtain confidence intervals on $G$ and assess whether the observed value is significantly different from the null expectation.
\end{enumerate}
For typical high-resolution Duffing simulations with $N \ge 10^4$ samples, the asymptotic $\chi^2$ approximation is generally adequate, but for short sequences ($N < 500$) or highly imbalanced partitions, permutation or bootstrap approaches are more robust.
\end{remark}

\subsection{Chapman--Kolmogorov consistency}

For a time-homogeneous Markov chain with one-step matrix $\mathbf{P}$, the Chapman--Kolmogorov
equation requires $\mathbf{P}^{(2)}=\mathbf{P}^2$ for the two-step transition matrix.
To test this hypothesis, empirically estimate $\widehat{\mathbf{P}}$ (one-step) and
$\widehat{\mathbf{P}}^{(2)}$ (two-step) from the symbol sequence and compute the
Frobenius-norm discrepancy:
\[
\Delta_{\mathrm{CK}}=\norm{\widehat{\mathbf{P}}^{(2)}-\widehat{\mathbf{P}}^2}_F.
\]

\paragraph{Bootstrap hypothesis test.}
To assess whether observed discrepancy is statistically significant, we perform a
bootstrap test under $H_0$: the process is first-order Markov. The procedure is summarized in Table~\ref{tab:bootstrap_ck_test}.

\begin{table}[htbp]
\centering
\caption{Bootstrap hypothesis test for Chapman--Kolmogorov consistency}
\label{tab:bootstrap_ck_test}
\begin{tabular}{ll}
\toprule
\textbf{Step} & \textbf{Description} \\
\midrule
1. Generate data & Generate $B$ (e.g., $B=1000$) synthetic sequences of length $N$ from a \\
& first-order Markov chain with transition matrix $\widehat{\mathbf{P}}$. \\[0.5ex]
2. Bootstrap loop & For each bootstrap replicate $b=1,\ldots,B$: \\
& \quad (a) Estimate $\widehat{\mathbf{P}}^{*,b}$ (one-step) and $\widehat{\mathbf{P}}^{(2),*,b}$ (two-step) from \\
& \quad\quad\, the synthetic sequence. \\
& \quad (b) Compute $\Delta_{\mathrm{CK}}^{*,b} = \norm{\widehat{\mathbf{P}}^{(2),*,b} - (\widehat{\mathbf{P}}^{*,b})^2}_F$. \\[0.5ex]
3. Compute $p$-value & Compute the empirical $p$-value: \\
& \quad $p_{\mathrm{CK}} = \frac{1 + \#\{b: \Delta_{\mathrm{CK}}^{*,b} \ge \Delta_{\mathrm{CK}}^{\mathrm{obs}}\}}{B+1}$. \\[0.5ex]
4. Confidence interval & Construct a $95\%$ confidence interval using bootstrap quantiles: \\
& \quad $\text{CI}_{95\%} = [\Delta_{\mathrm{CK}}^{*,(0.025B)}, \Delta_{\mathrm{CK}}^{*,(0.975B)}]$. \\[0.5ex]
5. Decision & Reject the first-order Markov hypothesis if $p_{\mathrm{CK}} < 0.05$. \\
\bottomrule
\end{tabular}
\end{table}

Large values of $\Delta_{\mathrm{CK}}$ relative to the bootstrap distribution, or
equivalently small $p_{\mathrm{CK}}$, indicate that a single time-homogeneous Markov
matrix is insufficient at the sampling scale, suggesting higher-order memory effects
or time-inhomogeneity. When rejection occurs, see Section~\ref{sec:markov_extensions}
for standard extensions (higher-order Markov, hidden Markov, semi-Markov).
However, Chapman--Kolmogorov consistency at a given order is necessary but not sufficient for embedding the empirical transition matrix into a continuous-time GKSL semigroup; the latter requires additional spectral conditions on the transition matrix.
When continuous-time embedding is uncertain or fails, the discrete-time Kraus representation remains a valid, model-free approach to ensure complete positivity and trace preservation.

\subsection{Run-length (dwell-time) statistics}

\begin{figure}[t]
\centering
\includegraphics[width=0.99\textwidth]{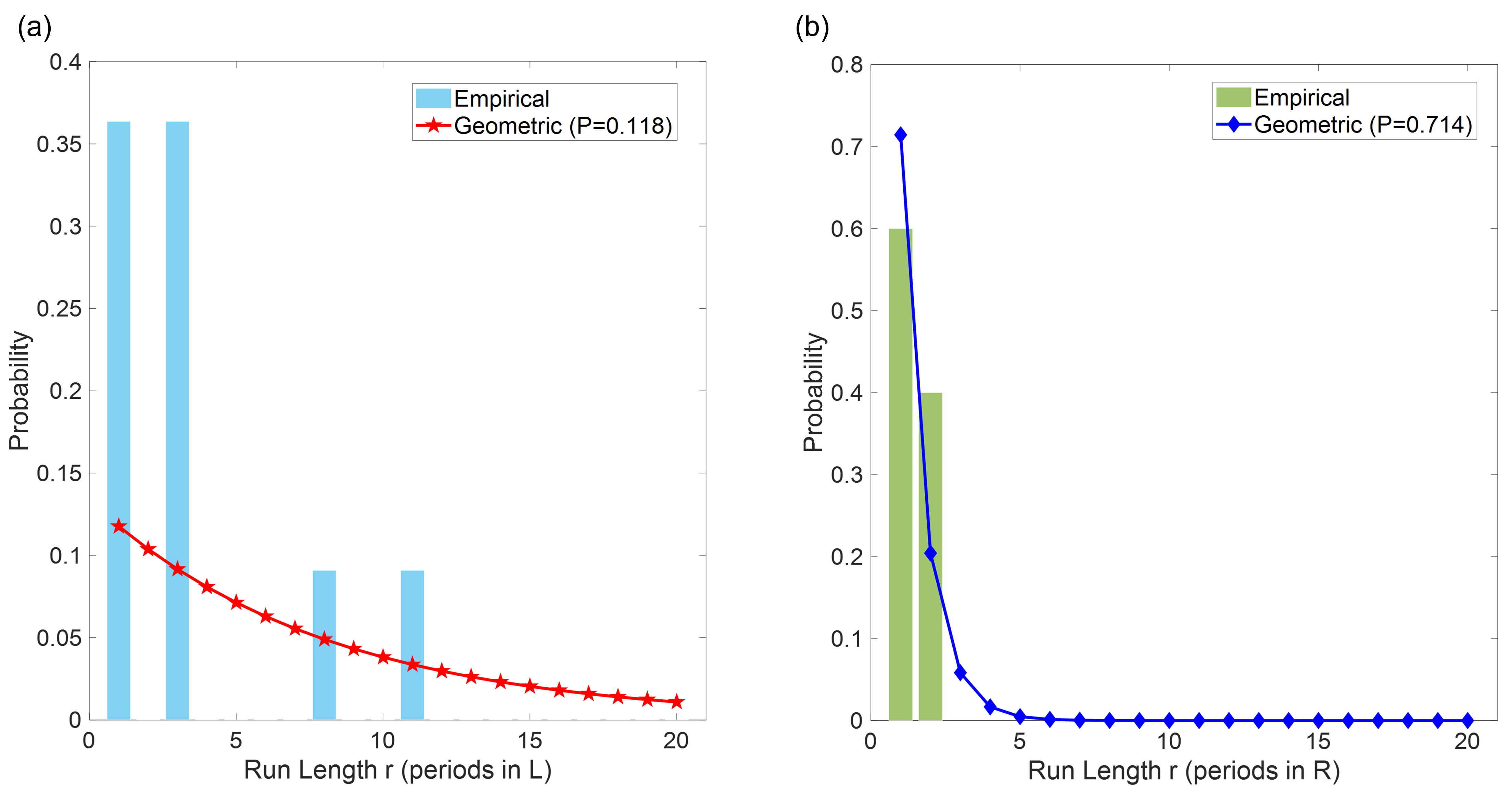}
\caption{Run-length (dwell-time) distributions testing the memoryless property of first-order Markov chains.
(a) Left-well run lengths: empirical histogram (blue bars) compared to geometric distribution $\Pr(R_L=r)=(1-P_{LR})^{r-1}P_{LR}$ (red curve) with transition probability $\hat P_{LR}$ estimated from data.
(b) Right-well run lengths: empirical histogram (orange bars) compared to geometric distribution $\Pr(R_R=r)=(1-P_{RL})^{r-1}P_{RL}$ (blue curve) with transition probability $\hat P_{RL}$ estimated from data.
Deviations from geometric laws diagnose non-Markov or time-inhomogeneous behavior at the sampling scale.
Parameters: $\alpha=1.0$, $\beta=1.0$, $\delta=0.15$, $\gamma_0=0.3$, $\omega=1.0$.}
\label{fig:runlengths}
\end{figure}

If $\{S_n\}$ is Markov with one-step probability $P_{LR}$, then the run length $R_L$ of consecutive $L$'s is geometric:
\[
\Pr(R_L=r)=(1-P_{LR})^{r-1}P_{LR},\qquad r\ge1,
\]
and similarly for $R$-runs.
Comparing empirical run-length histograms to these geometric laws is a direct test of memorylessness.
Figure~\ref{fig:runlengths} presents run-length distributions for both wells.

\begin{remark}[Well-dependent dwell-time behavior]
Experimental observations in chaotic bistable systems often reveal asymmetric run-length statistics.
For instance, the left well may exhibit predominantly short runs (1--3 periods) while the right well displays a broader distribution (1--15 periods).
Such well-dependent behavior can arise from asymmetric potential structure or differences in local escape mechanisms, and typically indicates deviations from a simple geometric law.
These deviations point to either higher-order memory or non-geometric (semi-Markov) waiting-time distributions.
\end{remark}

\subsection{Windowed stationarity test}

Split the post-transient sequence into $W$ windows (typically $W=5$ to $10$) of equal length and compute the transition probabilities $\hat P^{(w)}_{LR}$, $\hat P^{(w)}_{RL}$ for each window $w=1,\dots,W$.
Systematic drift beyond sampling variability suggests nonstationarity or time-dependent effective rates.

\paragraph{Quantitative threshold criteria.}
Three complementary approaches quantify whether observed window-to-window variation exceeds chance:
\begin{enumerate}[label=(\alph*), leftmargin=2em]
\item \textbf{Confidence-interval overlap:} Compute $95\%$ Wald confidence intervals for each $\hat P^{(w)}_{LR}$ (using $\mathrm{SE} = \sqrt{\hat P_{LR}(1-\hat P_{LR})/N_w}$, where $N_w$ is the number of L states in window $w$). Reject stationarity if intervals from different windows fail to overlap.
\item \textbf{Heterogeneity test:} Perform a $\chi^2$ test for homogeneity across windows: compute $\chi^2 = \sum_{w=1}^W \frac{(\hat P^{(w)}_{LR} - \bar P_{LR})^2}{\mathrm{Var}[\hat P^{(w)}_{LR}]}$, where $\bar P_{LR}$ is the pooled estimate. Under stationarity, $\chi^2 \sim \chi^2_{W-1}$.
\item \textbf{Change-point detection:} Apply a standard change-point algorithm (e.g., binary segmentation or PELT \cite{killick2012optimal}) to the time series $\{\hat P^{(w)}_{LR}\}_{w=1}^W$. Reject stationarity if a change point is detected at significance level $\alpha=0.05$.
\end{enumerate}
For typical Duffing simulations with steady forcing, confidence-interval overlap (approach a) is simplest and most transparent.
If nonstationarity is detected, a time-dependent generator $k_{LR}(t)$, $k_{RL}(t)$ or a richer coarse-graining (e.g., phase-resolved partitions) is required.

\subsection{If tests fail: controlled extensions}\label{sec:markov_extensions}

If the diagnostics reject the first-order time-homogeneous model, standard extensions are available.
First, one may adopt \textbf{higher-order Markov} models by augmenting the state space to words (e.g.\ $\{LL,LR,RL,RR\}$) so that the process becomes first-order in the enlarged space; for a second-order Markov model, the GKSL framework can be extended to a $4\times 4$ density matrix acting on the enlarged Hilbert space $\mathcal{H}_{2\text{-step}}=\mathrm{span}\{\ket{LL},\ket{LR},\ket{RL},\ket{RR}\}$, with jump operators encoding transitions among these four states.
Second, \textbf{hidden Markov} models may include a latent variable correlated with switching (e.g.\ drive phase or energy bin).
Third, \textbf{semi-Markov} approaches explicitly model non-geometric run lengths (non-exponential waiting times).
Fourth, one may allow \textbf{time-dependent rates} by fitting piecewise-constant or periodically modulated rate functions.
These modifications preserve the core ``two-lobe coarse-graining'' idea while making memory and nonstationarity explicit.

\begin{remark}[Experimental evidence of second-order effects]
High-resolution experimental data from deterministic Duffing simulations may reveal statistically significant second-order memory effects.
For example, the order test (Section~\ref{sec:markov_tests}) may yield a G-statistic with $p\ll 0.01$, rejecting the first-order null hypothesis.
Such findings do \emph{not} invalidate the GKSL framework; they indicate that the coarse-grained process at the chosen sampling scale is better described by a higher-order Markov model or a hidden Markov model.
Importantly, the Chapman--Kolmogorov consistency test (Section~\ref{sec:markov_tests}) can still hold, confirming that the Markov property is satisfied at a higher order and that the framework remains applicable with appropriate state-space augmentation.
\end{remark}

\begin{remark}[Caveat: not all discrete-time processes embed into continuous-time generators]
\label{rem:embedding_caveat}
While augmenting the state space to words allows construction of first-order Markov models in the enlarged space, not every such discrete-time transition matrix admits a representation as $\mathbf{P}=e^{\mathbf{Q}T}$ for a time-independent GKSL generator $\mathbf{Q}$.
This is the classical \emph{Markov embedding problem}: empirical transition matrices may violate necessary spectral conditions for continuous-time embedding.

\paragraph{Practical embedding diagnostics.}
For a two-state system, the embedding problem is usually tractable: a stochastic matrix $\mathbf{P}$ can be embedded in continuous time if and only if its eigenvalues are positive and the off-diagonal rates derived from the matrix logarithm are nonnegative.
For higher-dimensional systems (e.g., four-state second-order extensions), the following checks are recommended:
\begin{enumerate}[label=(\alph*), leftmargin=2em]
\item \textbf{Eigenvalue positivity:} Verify that all eigenvalues of $\widehat{\mathbf{P}}$ are strictly positive. Negative or complex eigenvalues with negative real parts preclude continuous-time embedding.
\item \textbf{Matrix logarithm and rate extraction:} Compute $\mathbf{Q} = \frac{1}{T}\log\widehat{\mathbf{P}}$ (using matrix logarithm). Check that the off-diagonal entries of $\mathbf{Q}$ (interpreted as jump rates) are nonnegative and that diagonal entries satisfy $Q_{ii} = -\sum_{j \ne i} Q_{ij}$. If rates are negative or the semigroup constraint fails, continuous embedding does not exist.
\item \textbf{Semigroup consistency:} Verify Chapman--Kolmogorov consistency at multiple time steps: check that $\widehat{\mathbf{P}}^{(kT)} \approx (\widehat{\mathbf{P}}^{(T)})^k$ for $k=2,3,\dots$. Large deviations suggest time-inhomogeneity or non-semigroup structure, ruling out a time-independent GKSL generator.
\end{enumerate}
When these diagnostics fail, we recommend working directly with the discrete-time CPTP map (Kraus representation) as the fundamental object, rather than attempting to infer a continuous-time GKSL generator.
Continuous-time GKSL generators should only be inferred when the data structure and Chapman--Kolmogorov tests strongly support time-homogeneous semigroup closure.
For two-state systems with typical Duffing parameters, continuous embedding usually succeeds; for higher-order extensions or systems with complex memory structure, discrete-time Kraus operators are often more appropriate.
\end{remark}

\subsection{Kraus representation and equivalence to generalized amplitude damping}\label{sec:kraus}

The continuous-time GKSL evolution $\frac{\dd\rho}{\dd t} = \mathcal{L}[\rho]$ can be integrated to yield a discrete-time completely positive trace-preserving (CPTP) map $\rho(t+T) = \Phi_T[\rho(t)]$.
For finite-time propagation, this map admits a Kraus (operator-sum) representation, which we now derive explicitly and prove equivalence to the well-known generalized amplitude damping (GAD) channel composed with dephasing and phase rotation.

\paragraph{Notation.} In this section we write $\rho$ for the model state $\rho_{\mathrm{mod}}$ to simplify notation in the Kraus-operator expressions.

Beyond the differential equation form, it is often useful to express the evolution as an operator-sum (Kraus) representation
\[
\rho(t)=\Lambda_t(\rho(0))=\sum_{\ell}K_\ell(t)\rho(0)K_\ell(t)^\dagger,\qquad
\sum_\ell K_\ell(t)^\dagger K_\ell(t)=\id,
\]
which makes complete positivity and trace preservation explicit.

For the two-rate model in Definition~\ref{def:model}, one can construct a Kraus family by composing:
(i) a generalized amplitude-damping channel encoding population relaxation toward $\rho^\infty$,
(ii) a phase-damping channel encoding dephasing with rate $\kappa$, and
(iii) a unitary phase rotation generated by $H$.

\subsection{Kraus operators}

Define
\[
\Gamma=k_{LR}+k_{RL},\qquad
\lambda(t)=1-e^{-\Gamma t},\qquad
\eta(t)=e^{-2\kappa t},
\qquad
p_R^\infty=\frac{k_{LR}}{\Gamma},\ p_L^\infty=\frac{k_{RL}}{\Gamma}.
\]
Let
\[
U(t)=e^{-iHt/\hbar}=\exp\!\left(+i\frac{\Omega t}{2}\sigma_z\right)
=
\begin{pmatrix}
e^{+i\Omega t/2} & 0\\
0 & e^{-i\Omega t/2}
\end{pmatrix}.
\]
Introduce generalized-amplitude-damping Kraus operators
\begin{align}
E_0(t) &= \sqrt{p_L^\infty}\begin{pmatrix} 1 & 0 \\ 0 & \sqrt{1-\lambda(t)} \end{pmatrix}, &
E_1(t) &= \sqrt{p_L^\infty}\begin{pmatrix} 0 & \sqrt{\lambda(t)} \\ 0 & 0 \end{pmatrix},\label{eq:E01}\\
E_2(t) &= \sqrt{p_R^\infty}\begin{pmatrix} \sqrt{1-\lambda(t)} & 0 \\ 0 & 1 \end{pmatrix}, &
E_3(t) &= \sqrt{p_R^\infty}\begin{pmatrix} 0 & 0 \\ \sqrt{\lambda(t)} & 0 \end{pmatrix},\label{eq:E23}
\end{align}
and phase-damping Kraus operators
\begin{equation}\label{eq:Fa}
F_0(t)=\sqrt{\frac{1+\eta(t)}{2}}\,\id,\qquad
F_1(t)=\sqrt{\frac{1-\eta(t)}{2}}\,\sigma_z.
\end{equation}
Define the composite family
\begin{equation}\label{eq:M_aj_def}
M_{aj}(t)=U(t)\,F_a(t)\,E_j(t),\qquad a\in\{0,1\},\ j\in\{0,1,2,3\}.
\end{equation}

\begin{lemma}[Completeness]\label{lem:kraus_complete}
The operators $M_{aj}(t)$ satisfy $\sum_{a=0}^1\sum_{j=0}^3 M_{aj}(t)^\dagger M_{aj}(t)=\id$.
\end{lemma}

\begin{remark}
Lemma~\ref{lem:kraus_complete} implies that $\Lambda_t(\rho)=\sum_{a,j}M_{aj}(t)\rho M_{aj}(t)^\dagger$ is CPTP.
A direct calculation also shows that this map reproduces the closed-form GKSL solutions for populations and off-diagonals; see Theorem~\ref{thm:kraus_equivalence} and Appendix~\ref{app:kraus_proof}.
\end{remark}

\begin{theorem}[Equivalence to the GKSL solution]\label{thm:kraus_equivalence}
Let $\Lambda_t(\rho)=\sum_{a,j}M_{aj}(t)\rho M_{aj}(t)^\dagger$ with \eqref{eq:M_aj_def}.
Then for any initial state $\rho(0)$,
\begin{align}
\rho_{LL}(t) &= p_L^\infty+(\rho_{LL}(0)-p_L^\infty)e^{-\Gamma t},\label{eq:kraus_pop}\\
\rho_{LR}(t) &= \rho_{LR}(0)\exp\!\left[\left(i\Omega-\frac{\Gamma}{2}-2\kappa\right)t\right].\label{eq:kraus_coh}
\end{align}
Hence $\Lambda_t$ coincides with the GKSL evolution generated by Definition~\ref{def:model}.
\end{theorem}

\begin{remark}[Interpretational note]
The Kraus representation is a statement about the \emph{reduced two-state model}.
It does not imply that the underlying Duffing dynamics is quantum; rather, it provides a mathematically consistent CPTP parametrization for the coarse-grained statistics.
\end{remark}

\section{Numerical methods}\label{sec:numerics_pipeline}

This section outlines computational validation procedures for applying the framework to Duffing or other bistable systems.
We provide a concise protocol for integration, symbol generation, diagnostic testing, and rate estimation to facilitate reproducible numerical studies.

\subsection{Simulation and sampling protocol}

Integrate the driven Duffing system \eqref{eq:duffing} using a stable ODE solver (e.g., 4th-order Runge--Kutta or adaptive Dormand--Prince) with time step $\Delta t \ll T$ where $T=2\pi/\omega$ is the forcing period.
For stochastic extensions \eqref{eq:duffing_noise} with noise intensity $D>0$, use Euler--Maruyama or higher-order SDE schemes.
Discard initial transients (typically 50--100 periods) and collect $N$ Poincar\'e samples at times $t_n = nT$ over 200--500 periods to ensure adequate statistics for diagnostic tests.
Verify numerical convergence by halving $\Delta t$ and checking that transition probabilities change by less than 1\%.

\subsection{Symbol generation and soft memberships}

Construct hard symbols $\{S_n\}$ using Definition~\ref{def:symbol} (sign-based partition) or a refined boundary function $g(z)$ based on approximate separatrix geometry.
For soft-partition analysis, compute membership weights $w_L(z_n)$, $w_R(z_n)$ via Definition~\ref{def:soft_partition} for a grid of fuzziness parameters $\varepsilon \in [10^{-2}, 10^1]$.
Estimate the embedded state $\rho_{\mathrm{emb}}$ and overlap parameter $c(\varepsilon)$ via Proposition~\ref{prop:rho_elements_overlap}, and plot $c(\varepsilon)$ vs.\ $\varepsilon$ to verify the hard-partition limit $c(\varepsilon) \to 0$ as $\varepsilon \to 0^+$ (Figure~\ref{fig:c_vs_epsilon}).

\subsection{Diagnostic testing}

Apply the four tests developed in Section~\ref{sec:markov_tests} to the symbol sequence $\{S_n\}$:
\begin{enumerate}[label=(\alph*), leftmargin=2em]
\item \textbf{Order test}: Compute G-statistic via \eqref{eq:g_test} and report $p$-value; reject first-order Markov if $p < 0.05$.
\item \textbf{Chapman--Kolmogorov}: Estimate $\widehat{\mathbf{P}}$ and $\widehat{\mathbf{P}}^{(2)}$, compute discrepancy $\Delta_{\mathrm{CK}} = \|\widehat{\mathbf{P}}^{(2)} - \widehat{\mathbf{P}}^2\|_F$, and obtain bootstrap confidence intervals using the procedure in Table~\ref{tab:bootstrap_ck_test}.
\item \textbf{Run-length distributions}: Plot histograms of run lengths for L and R states, overlay geometric fits, and assess goodness-of-fit via $\chi^2$ test.
\item \textbf{Windowed stationarity}: Divide the steady-state sequence into $W=5$ to $10$ windows, estimate $\hat P_{LR}^{(w)}$ for each window, and test for heterogeneity using confidence-interval overlap or $\chi^2$ test.
\end{enumerate}
If diagnostics reject first-order Markov closure, apply the state-space augmentation protocols outlined in Section~\ref{sec:markov_extensions}.

\subsection{Rate estimation and model validation}

If Markov diagnostics pass, estimate switching rates via the inversion formulas \eqref{eq:Gamma_hat} and \eqref{eq:k_hat}.
Compare model-predicted populations $p_L(t)$ (from solving \eqref{eq:rhoLL}--\eqref{eq:rhoRR} with estimated rates) against empirical populations from windowed time averages.
Quantify agreement using mean absolute error or Kullback--Leibler divergence.
Report estimated rates $\hat k_{LR}$, $\hat k_{RL}$, steady-state populations $p_L^\infty$, and relaxation time $\tau = 1/\hat\Gamma$ with confidence intervals obtained via block bootstrap resampling (Table~\ref{tab:block_bootstrap}).

\section{Discussion}\label{sec:numerical_discussion}

\paragraph{Soft memberships.}
Fix $g(z)$ and a fuzziness parameter $\varepsilon$, then compute $w_L(z(t_n)),w_R(z(t_n))$ by \eqref{eq:soft_wL}.
Estimate
\begin{equation}\label{eq:c_hat}
\hat c(\varepsilon)=\frac{1}{N}\sum_{n=1}^N \sqrt{w_L(z(t_n))w_R(z(t_n))}.
\end{equation}
Repeat for a grid of $\varepsilon$ values to verify $\hat c(\varepsilon)\to0$ as $\varepsilon\to0^+$.

\subsection{Estimating transition probabilities and continuous-time rates}

Compute the transition counts
\[
N_{LL},N_{LR},N_{RL},N_{RR},
\]
and the empirical one-step probabilities $\hat P_{LR},\hat P_{RL}$.
Then reconstruct $\hat\Gamma,\hat k_{LR},\hat k_{RL}$ using \eqref{eq:Gamma_hat}--\eqref{eq:k_hat}.
Report confidence intervals, e.g.\ by binomial standard errors or block bootstrap (Table~\ref{tab:block_bootstrap}; the latter is preferred if serial correlations are strong).

\subsection{Block bootstrap for confidence intervals on transition probabilities}\label{sec:block_bootstrap}

When serial correlation is strong in the symbol sequence (e.g., due to quasi-periodic
dynamics or memory effects), standard binomial confidence intervals for estimated
transition probabilities $\hat{P}_{LR}$ and $\hat{P}_{RL}$ are too narrow. We employ
the block bootstrap \cite{politis1994,lahiri1999} to obtain robust confidence intervals.

\paragraph{Block length selection.}
The optimal block length $\ell$ balances bias (too short blocks lose dependency structure) and variance (too long blocks reduce effective sample size).
A practical choice is
\begin{equation}\label{eq:block_length}
\ell \approx 2\tau_{\mathrm{int}},
\end{equation}
where $\tau_{\mathrm{int}}$ is the integrated autocorrelation time of the indicator process $I[S_n=L]$, estimated via
\[
\hat\tau_{\mathrm{int}} = 1 + 2\sum_{k=1}^{K_{\max}} \hat\rho(k),
\]
with $\hat\rho(k)$ the sample autocorrelation function truncated at a suitable lag $K_{\max}$ (e.g., first zero crossing or $K_{\max} = \lfloor N^{1/3}\rfloor$).
Alternatively, overlapping (moving) blocks can increase the number of distinct resampled blocks; the stationary bootstrap \cite{politis1994} provides an automatic adaptive procedure.

\paragraph{Block bootstrap algorithm.}
The block bootstrap procedure for computing confidence intervals on transition probabilities and rate parameters is detailed in Table~\ref{tab:block_bootstrap}.

\begin{table}[htbp]
\centering
\caption{Block bootstrap algorithm for confidence intervals on transition probabilities}
\label{tab:block_bootstrap}
\begin{tabular}{ll}
\toprule
\textbf{Step} & \textbf{Description} \\
\midrule
1. Block division & Divide the post-transient sequence $\{S_1,\ldots,S_N\}$ into $m=\lfloor N/\ell\rfloor$ \\
& contiguous, non-overlapping blocks of length $\ell$. \\[0.5ex]
2. Bootstrap loop & For $b=1,\ldots,B$ bootstrap replicates (e.g., $B=1000$): \\
& \quad (a) Randomly draw $m$ blocks with replacement from the $m$ available blocks. \\
& \quad (b) Concatenate to form a bootstrap sequence $\{S_1^*,\ldots,S_N^*\}$. \\
& \quad (c) Compute bootstrap estimates: $\hat{P}_{LR}^{*,b}$, $\hat{P}_{RL}^{*,b}$, and \\
& \quad\quad\, (if desired) $\hat{k}_{LR}^{*,b}$, $\hat{k}_{RL}^{*,b}$ using Eqs.~\eqref{eq:Gamma_hat}--\eqref{eq:k_hat}. \\[0.5ex]
3. Confidence intervals & Construct $95\%$ confidence intervals using the percentile method: \\
& \quad $\text{CI}_{95\%} = \left[\hat{\theta}^{*,(0.025B)}, \hat{\theta}^{*,(0.975B)}\right]$, \\
& where $\hat{\theta}^{*,(q)}$ denotes the $q$-th empirical quantile of the bootstrap \\
& replicates $\{\hat{\theta}^{*,1},\ldots,\hat{\theta}^{*,B}\}$. \hfill (Eq.~\ref{eq:block_bootstrap_ci}) \\[0.5ex]
4. Standard error & Compute block-bootstrap standard error: \\
& \quad $\widehat{\mathrm{SE}}_{\mathrm{block}} = \sqrt{\frac{1}{B}\sum_{b=1}^B\left(\hat{\theta}^{*,b} - \bar{\theta}^*\right)^2}$, \\
& where $\bar{\theta}^* = \frac{1}{B}\sum_{b=1}^B \hat{\theta}^{*,b}$. \hfill (Eq.~\ref{eq:block_se}) \\
\bottomrule
\end{tabular}
\end{table}

\begin{equation}\label{eq:block_bootstrap_ci}
\text{CI}_{95\%} = \left[\hat{\theta}^{*,(0.025B)}, \hat{\theta}^{*,(0.975B)}\right]
\end{equation}

\begin{equation}\label{eq:block_se}
\widehat{\mathrm{SE}}_{\mathrm{block}} = \sqrt{\frac{1}{B}\sum_{b=1}^B\left(\hat{\theta}^{*,b}
- \bar{\theta}^*\right)^2}, \quad
\bar{\theta}^* = \frac{1}{B}\sum_{b=1}^B \hat{\theta}^{*,b}
\end{equation}

\paragraph{Reporting.}
When presenting estimated rates, report the block-bootstrap point estimate and
$95\%$ confidence interval. For example:
\[
\hat{k}_{LR} = 0.42 \text{ (95\% CI: } [0.38, 0.46]).
\]
This acknowledges serial correlation and provides a more honest quantification of
uncertainty than standard errors assuming independence.

\subsection{Model checks (Markov diagnostics)}

Apply the diagnostics in Section~\ref{sec:markov_tests}.
Specifically, perform the order test via \eqref{eq:g_test}, compute the Chapman--Kolmogorov discrepancy $\Delta_{\mathrm{CK}}$, compare the run-length histogram against the geometric law, and assess windowed stationarity of $\hat P_{LR},\hat P_{RL}$.
If diagnostics fail, explicitly switch to an extended model and report the extension (higher-order, hidden Markov, semi-Markov, or time-dependent rates).

\subsection{Numerical results}

\begin{figure}[t]
\centering
\includegraphics[width=0.99\textwidth]{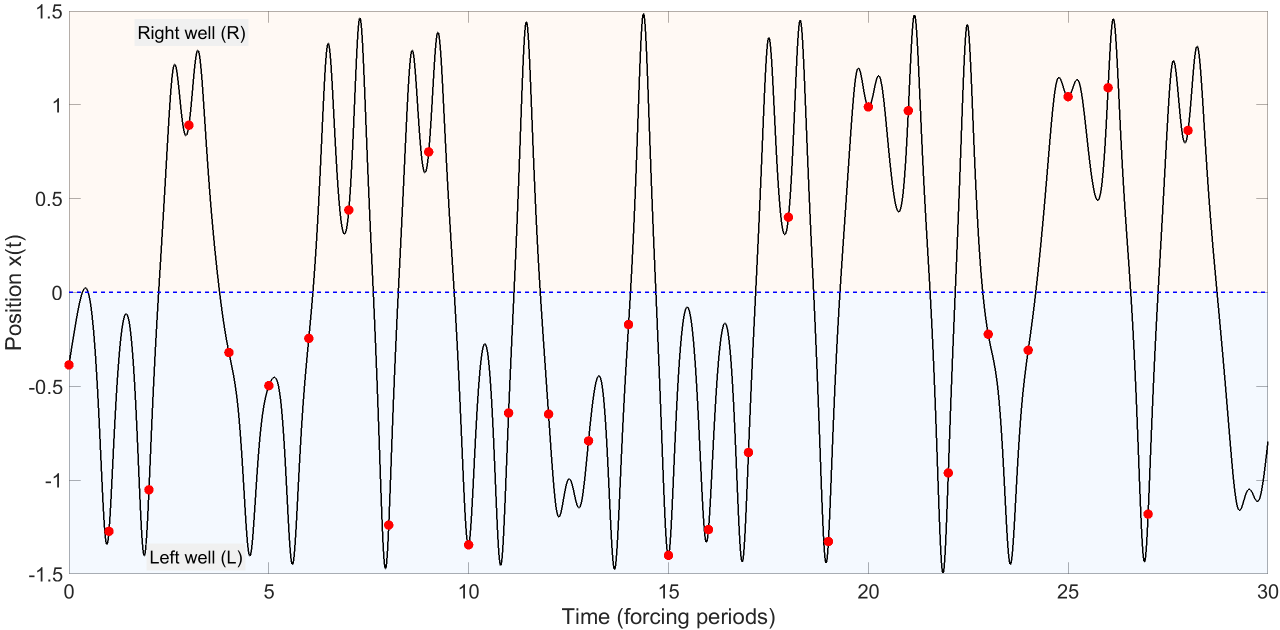}
\caption{Representative time series $x(t)$ in a chaotic bistable regime, with the two wells indicated (e.g.\\ $x<0$ and $x>0$ regions) and stroboscopic sampling points marked. This figure justifies that the chosen parameter set exhibits inter-well hopping on the observation horizon.}
\label{fig:duffing_timeseries}
\end{figure}

\begin{table}[t]
\centering
\caption{Transition counts and one-step probabilities at sampling interval $T=2\pi/\omega$.}
\label{tab:transition_counts}
\begin{tabular}{lcc}
\toprule
 & to $L$ & to $R$\\
\midrule
from $L$ & $75$ & $10$\\
from $R$ & $10$ & $4$\\
\midrule
$\hat P_{LR}$ & \multicolumn{2}{c}{$0.1176$}\\
$\hat P_{RL}$ & \multicolumn{2}{c}{$0.7143$}\\
\bottomrule
\end{tabular}
\end{table}

\begin{table}[t]
\centering
\caption{Reconstructed continuous-time rates from \eqref{eq:Gamma_hat}--\eqref{eq:k_hat}. Report uncertainty estimates.}
\label{tab:rates}
\begin{tabular}{lc}
\toprule
Quantity & Estimate\\
\midrule
$\hat\Gamma$ & $0.2838 \pm 0.1397$\\
$\hat k_{LR}$ & $0.2437 \pm 0.1245$\\
$\hat k_{RL}$ & $0.0401 \pm 0.0207$\\
\bottomrule
\end{tabular}
\end{table}

\begin{figure}[t]
\centering
\includegraphics[width=0.99\textwidth]{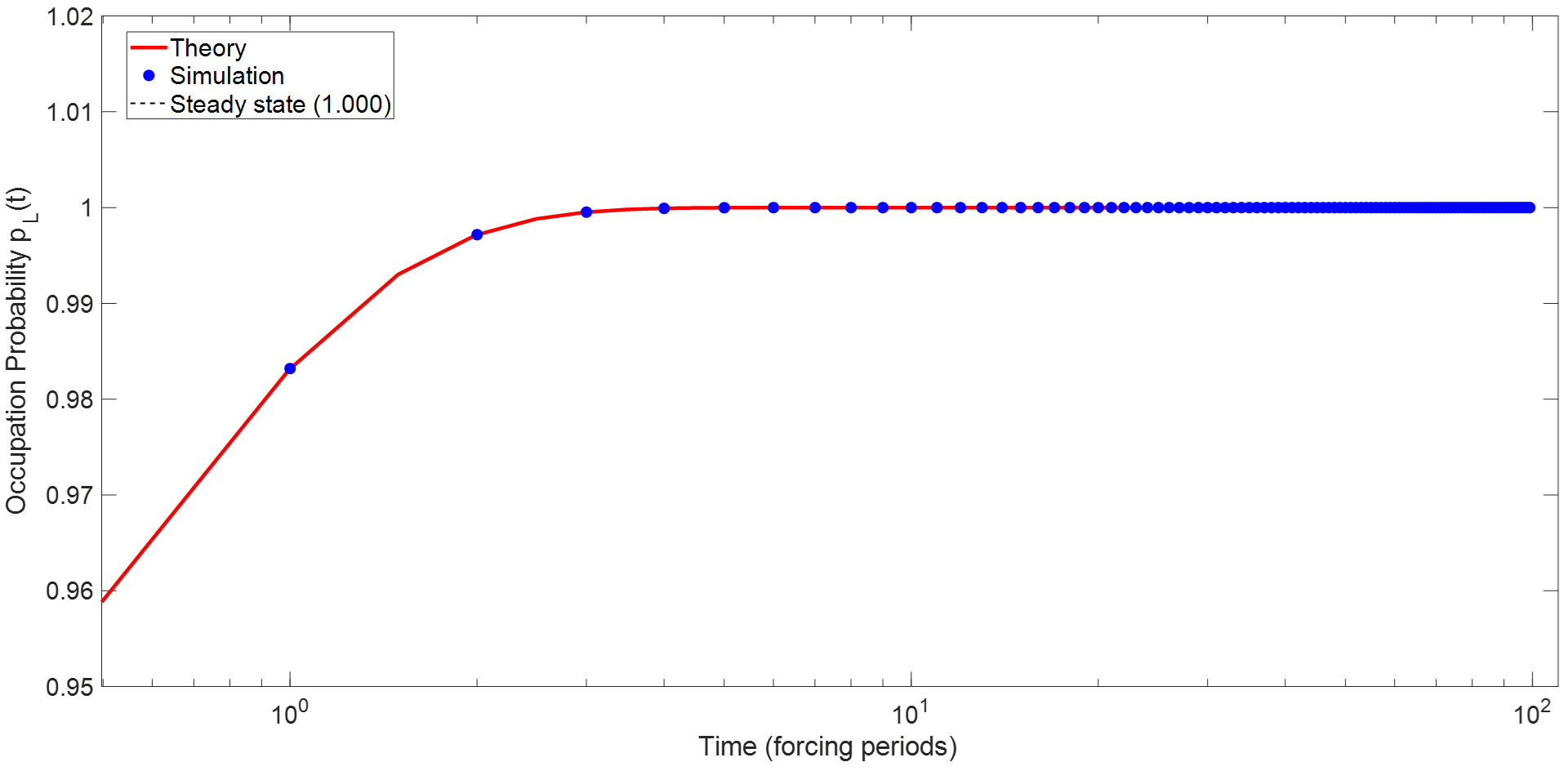}
\caption{Population relaxation: comparison of the theoretical curve $p_L(t)=p_L^\infty+(p_0-p_L^\infty)e^{-\Gamma t}$ (Theorem~\ref{thm:pop_solution}) to an ensemble-averaged estimate of $p_L(t)$ from simulation, started from biased initial conditions.}
\label{fig:population_relaxation}
\end{figure}

\begin{figure}[t]
\centering
\includegraphics[width=0.99\textwidth]{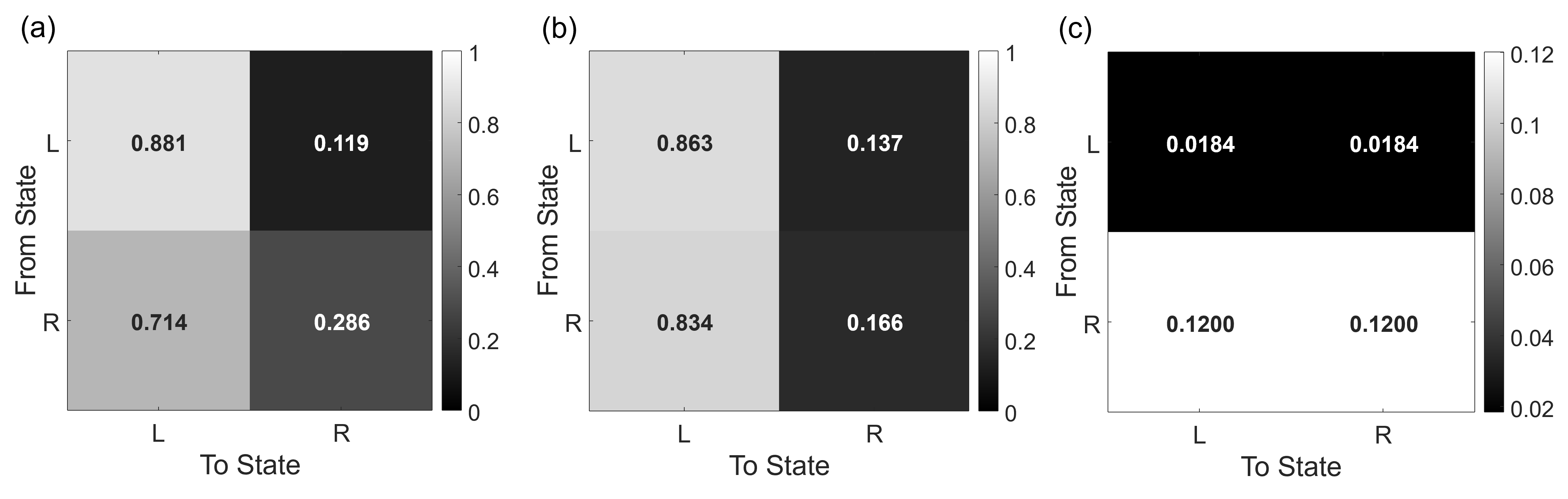}
\caption{Chapman--Kolmogorov consistency test comparing empirical and predicted two-step transition probabilities.
(a) Empirical two-step matrix $\widehat{\mathbf{P}}^{(2)}$ estimated from symbol pairs separated by two periods.
(b) Predicted two-step matrix $\widehat{\mathbf{P}}^2$ computed by squaring the one-step matrix.
(c) Element-wise discrepancy $|\widehat{\mathbf{P}}^{(2)}-\widehat{\mathbf{P}}^2|$ with Frobenius norm $\Delta_{\mathrm{CK}}$.
Small discrepancies indicate Chapman--Kolmogorov consistency, validating the Markov property across time scales (Section~\ref{sec:markov_tests}).
Parameters: $\alpha=1.0$, $\beta=1.0$, $\delta=0.15$, $\gamma_0=0.3$, $\omega=1.0$.}
\label{fig:ck_test}
\end{figure}

\begin{figure}[t]
\centering
\includegraphics[width=0.99\textwidth]{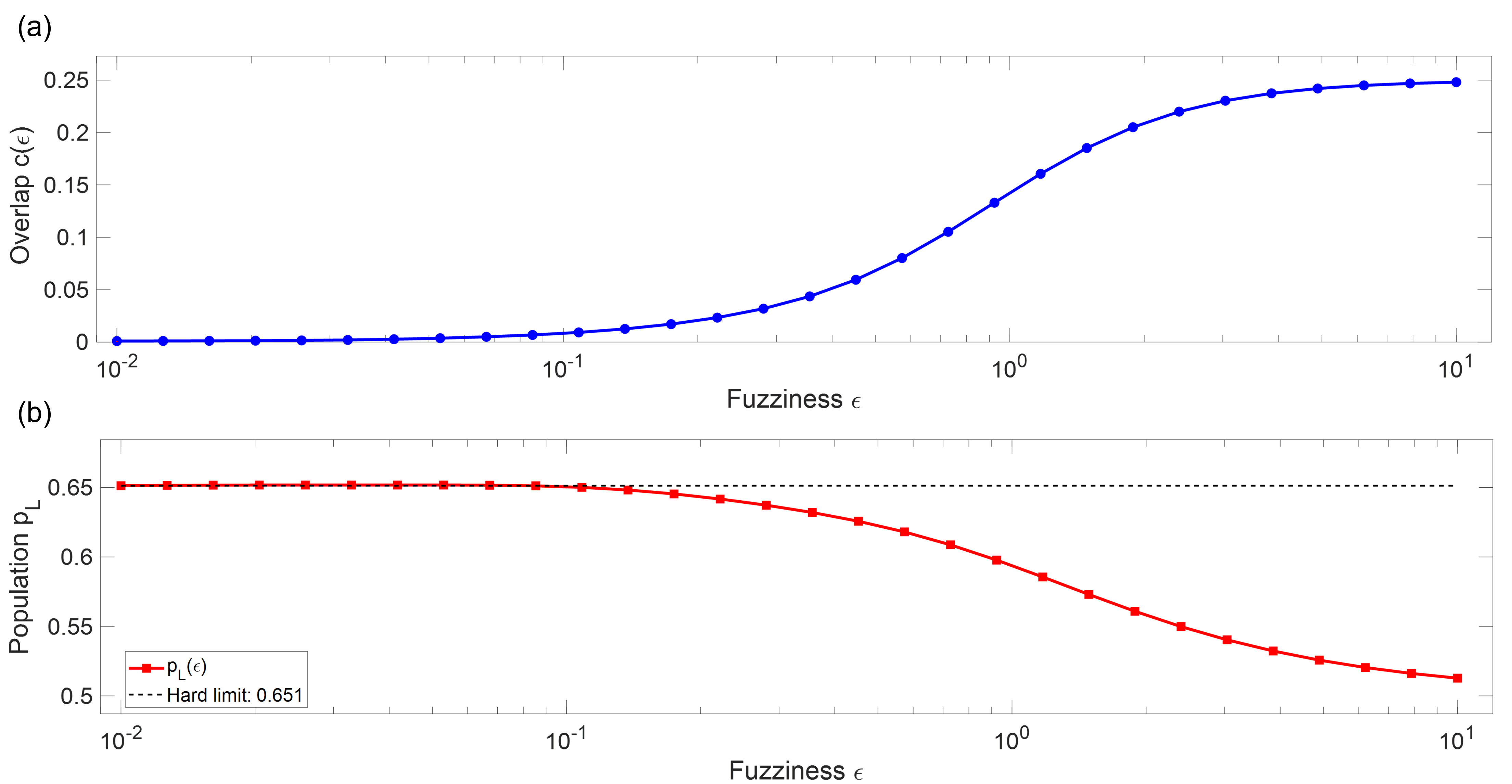}
\caption{Overlap parameter $\hat c(\varepsilon)$ and population $p_L(\varepsilon)$ versus soft-partition fuzziness $\varepsilon$.
(a) Overlap $\hat c(\varepsilon)=\langle \sqrt{w_L w_R}\rangle$ decreases toward zero as $\varepsilon\to 0^+$, demonstrating convergence to hard-partition limit (Proposition~\ref{prop:rho_elements_overlap}).
(b) Left-well population $p_L(\varepsilon)=\langle w_L\rangle$ converges to the hard-partition value as $\varepsilon\to 0^+$ and varies with increased fuzziness.
This quantifies the sensitivity of the embedded state $\rho_{\mathrm{emb}}(\varepsilon)$ to the resolution parameter $\varepsilon$ (Section~\ref{sec:coarse_graining}).
These diagnostics characterize how the embedding choice (controlled by $\varepsilon$) affects the coarse-grained state representation, but are \emph{not} predicted by the GKSL dynamics, which governs only population evolution from symbol observations (Section~\ref{sec:coarse_graining}).
The GKSL steady state is diagonal ($\rho_{LR}^\infty=0$, Theorem~\ref{thm:steady}) because symbols contain no boundary-proximity information.
Parameters: $\alpha=1.0$, $\beta=1.0$, $\delta=0.15$, $\gamma_0=0.3$, $\omega=1.0$.}
\label{fig:c_vs_epsilon}
\end{figure}

\subsection{Minimal reporting checklist}

To make the numerical study reproducible and the statistical claims falsifiable, we recommend reporting six key items.
First, provide full Duffing parameters $(\alpha,\beta,\delta,\gamma_0,\omega)$ and solver details (integrator, $\Delta t$, convergence checks).
Second, specify the sampling scheme (Poincar\'e period $T$, window length $\Delta$ if used).
Third, report total samples $N$ and transient discarded $N_\mathrm{trans}$.
Fourth, present transition counts and estimated $(\hat P_{LR},\hat P_{RL})$, together with reconstructed rates $(\hat k_{LR},\hat k_{RL})$ with uncertainty.
Fifth, document Markov diagnostic outcomes (order test statistic/p-value, CK discrepancy, run-length fits, stationarity results).
Sixth, if diagnostics fail, report the chosen extension and its fitted parameters.

\section{Discussion and limitations}\label{sec:discussion}

\subsection{What the framework provides}

At the level of diagonal states, the two-rate GKSL model is equivalent to the classical two-state master equation.
The value of the Lindblad--Pauli presentation is therefore not in changing the underlying stochastic assumption, but in \emph{organizing} it:
the GKSL generator makes CPTP structure explicit, the Kraus form provides a constructive channel representation, and the Bloch parametrization offers a geometric view of the reduced state space.

The soft-partition embedding adds an additional operational parameter $c(t)$ that captures boundary overlap of the coarse-grained distribution.
Even though $c(t)$ is not a microscopic quantum coherence, it is a well-defined functional of the chosen coarse-graining and can be estimated from data.
This may be useful when comparing different coarse-grainings or quantifying the effect of finite resolution/windowing on inferred switching statistics.

Crucially, $c(\varepsilon)$ is \emph{not} the same as the off-diagonal element $\rho_{LR}$ in the GKSL model.
The GKSL dynamics govern populations from symbol observations; off-diagonal terms in this model represent initial-state uncertainty and decay to zero at steady state (Theorem~\ref{thm:steady}).
In contrast, $c(\varepsilon)$ is computed directly from phase-space data and characterizes the chosen partition resolution.
Figure~\ref{fig:c_vs_epsilon} shows $c(\varepsilon)$ as a function of the fuzziness parameter, demonstrating that sharper partitions ($\varepsilon\to 0$) yield $c\to 0$ (classical limit), while soft partitions have $c(\varepsilon)>0$.
This diagnostic helps select an appropriate coarse-graining scale by balancing resolution (small $\varepsilon$, $c\approx 0$) against the need to avoid spurious boundary-crossing artifacts in chaotic trajectories.

\subsection{Limitations}

\paragraph{Model closure.}
A two-state Markov closure can fail if (i) switching is influenced by slow variables (energy, phase, or intra-well chaotic motion) that retain memory across sampling steps, (ii) the observation interval $T$ is not well separated from relevant correlation times, or (iii) the system exhibits nonstationary regimes.
Section~\ref{sec:markov_tests} provides diagnostics and standard extensions.

\paragraph{Interpretation of off-diagonal terms.}
The off-diagonal term $c(t)$ is embedding-dependent.
If only $L/R$ occupancy is measured, $c(t)$ cannot be inferred uniquely without additional operational definitions (boundary observables).
Thus $c(t)$ should be treated as a coarse-graining descriptor rather than as an intrinsic property of the underlying continuous dynamics.

\paragraph{Parameter dependence.}
The rates $k_{LR},k_{RL}$ are phenomenological in this framework.
Deriving them from first principles for deterministic chaotic switching is nontrivial and generally requires additional dynamical-systems analysis.
Our emphasis is on measurable inference and falsifiable reduced modeling.

\subsection{Model order and extensions}

Experimental validation of the first-order Markov assumption is critical.
High-resolution numerical experiments on deterministic Duffing systems can reveal statistically significant second-order memory effects, as evidenced by order tests (e.g., G-test with $p\approx 0.001$).
Such findings indicate that the coarse-grained symbol sequence $\{S_n\}$ at the chosen sampling interval retains memory of the previous two states rather than just the immediate predecessor.

\paragraph{Interpretation and framework validity.}
The presence of second-order memory does \emph{not} undermine the GKSL framework.
Rather, it signals that the effective state space should be augmented.
For a second-order Markov process, one can construct a $4\times 4$ density matrix on the Hilbert space $\mathcal{H}_{2\text{-step}}=\mathrm{span}\{\ket{LL},\ket{LR},\ket{RL},\ket{RR}\}$, and define jump operators $L_{ijkl}$ that map $(S_{n-1},S_n)\to(S_n,S_{n+1})$.
This enlarged GKSL model is first-order Markov in the augmented state and can capture the second-order dependencies observed experimentally.

\paragraph{Chapman--Kolmogorov as validation.}
Crucially, the Chapman--Kolmogorov consistency test can still hold even when first-order order tests fail.
This indicates that the process is Markov at a higher order, and that the GKSL formalism remains applicable with an appropriately enlarged state space.
The framework's value lies in its systematic approach to constructing completely positive trace-preserving (CPTP) maps, which remains valid regardless of the order of the Markov model.

\paragraph{Practical recommendations.}
When experimental diagnostics reject the first-order assumption, several extensions are available.
One may augment the state space to pairs or triplets of symbols (Section~\ref{sec:markov_extensions}).
Alternatively, one may consider hidden Markov models (HMMs) that include latent variables such as drive phase or energy level.
For systems with well-dependent dwell-time statistics, semi-Markov models with explicit waiting-time distributions may be more parsimonious.

\subsection{Extensions and model diagnostics}

Natural next steps include: multi-state coarse-grainings (e.g., adding a boundary state), time-dependent generators, semi-Markov dwell-time models, and systematic comparisons between hard- and soft-partition embeddings.
Each extension inherits the CPTP structure and remains amenable to the same diagnostic framework.

\paragraph{Quantum information tools for model diagnostics.}
Because the coarse-grained evolution satisfies CPTP properties (automatic from GKSL form), standard quantum-information quantities provide objective model-selection and validation metrics.
\textbf{Purity} $\mathrm{Tr}(\rho_{\mathrm{emb}}^2)$ quantifies the classicality of the effective state: high purity ($\to1$) signals clean binary switching, while reduced purity ($< 0.9$) indicates significant partition overlap or missing state-space degrees of freedom.
Plotting purity vs.\ coarse-graining resolution $\varepsilon$ identifies the resolution-vs.-purity trade-off.
\textbf{Fidelity} $F(\rho_{\mathrm{mod}}, \rho_{\mathrm{emb}}) = \mathrm{Tr}\sqrt{\sqrt{\rho_{\mathrm{emb}}}\,\rho_{\mathrm{mod}}\,\sqrt{\rho_{\mathrm{emb}}}}$ measures agreement between GKSL predictions and empirical densities; maximizing fidelity over parameters $(k_{LR}, k_{RL})$ performs nonparametric rate estimation, and fidelity thresholds (e.g., $F > 0.95$) validate model adequacy on held-out data.
\textbf{Trace distance} $D_{\mathrm{tr}}(\rho_{\mathrm{hard}}, \rho_{\mathrm{soft}}) = \tfrac12\,\mathrm{Tr}\lvert\rho_{\mathrm{hard}} - \rho_{\mathrm{soft}}\rvert$ assesses robustness to partition-width choices: small $D_{\mathrm{tr}}$ across a range of $\varepsilon$ indicates stable inferred rates, whereas large $D_{\mathrm{tr}}$ indicates partition sensitivity and suggests the need for higher-resolution coarse-grainings.

\paragraph{Data requirements.}
These metrics require the embedded state $\rho_{\mathrm{emb}}(t;\varepsilon)$ constructed from phase-space or soft-membership data (Section~\ref{sec:coarse_graining}).
If only the symbol sequence $\{S_n\}$ is available, the off-diagonal element $c(\varepsilon)$ cannot be measured, and diagnostic tools must rely on diagonal quantities: symbol entropy, likelihood-ratio tests, run-length distributions (Section~\ref{sec:markov_tests}), and windowed stationarity checks.
Purity and fidelity are most informative when phase-space data permit estimation of both populations and boundary overlap.

These tools are classical operational quantities, applied to classical measurements of coarse-grained occupancy statistics.
They do not require quantum hardware; rather, they provide rigorous, interpretable diagnostics for model selection and uncertainty quantification.

\paragraph{Possible future roles for quantum-assisted analysis.}
If and when practical quantum processors become available for small systems (tens to hundreds of qubits), the ability to directly implement Lindblad propagators via jump-operator circuits offers an \emph{alternative simulation pathway}.
This would not provide computational speedup (classical matrix exponentiation is efficient for 2-state systems); rather, it would enable cross-platform validation of the model and exploration of very-high-dimensional extensions ($d \gg 10$) on quantum hardware.
For the current focus (2-state and $d < 10$ systems), classical methods suffice.

\section{Conclusion}\label{sec:conclusion}

We presented a Lindblad--Pauli framework for coarse-grained chaotic bistable dynamics, centered on a two-state embedding of Duffing left/right statistics into a $2\times2$ density matrix and a two-rate GKSL generator for inter-well switching.
The framework yields closed-form solutions, an explicit Kraus representation, and a set of falsifiable diagnostics for the Markov assumptions implicit in a two-state reduction.
A soft-partition embedding induces a Bloch half-disk state space and a coarse-graining diagnostic $c(\varepsilon)$ that quantifies partition fuzziness; this parameter characterizes the embedding geometry but is not a dynamical variable of the reduced model, which governs only population evolution from symbol observations.

When experimental diagnostics reveal deviations from first-order Markov behavior---such as second-order memory effects evidenced by G-test statistics with $p\ll 0.01$---the framework extends naturally through state-space augmentation to $4\times 4$ (or higher-dimensional) GKSL models.
Importantly, the Chapman--Kolmogorov consistency test can still hold in such cases, confirming that the Markov property is satisfied at a higher order and that the GKSL formalism remains valid and applicable.
This demonstrates that the framework is not rigidly tied to first-order models but provides a systematic, extensible approach to modeling coarse-grained dynamics with complete positivity guarantees.

A detailed numerical pipeline was provided to support computational validation.

\clearpage
\bibliographystyle{unsrt}
\bibliography{references}

\clearpage
\appendix
\section*{Appendices}
\addcontentsline{toc}{section}{Appendices}

\section{Detailed derivation of the evolution equations}\label{app:evolution_derivation}

\paragraph{Notation.} In this appendix we write $\rho$ for $\rho_{\mathrm{mod}}$ to simplify notation.

This appendix provides a direct derivation of Theorem~\ref{thm:evolution} for completeness.

Let $\rho=\begin{pmatrix}\rho_{LL}&\rho_{LR}\\\rho_{RL}&\rho_{RR}\end{pmatrix}$.
We evaluate the GKSL terms in \eqref{eq:lindblad_general} for the operators \eqref{eq:H_model}--\eqref{eq:L_d}.

\paragraph{Hamiltonian term.}
With $H=-(\Delta E/2)\sigma_z$,
\[
[H,\rho]=H\rho-\rho H=
\begin{pmatrix}
0 & -\Delta E\,\rho_{LR}\\
+\Delta E\,\rho_{RL} & 0
\end{pmatrix},
\]
hence
\[
-\frac{i}{\hbar}[H,\rho]=
\begin{pmatrix}
0 & +i\Omega\,\rho_{LR}\\
-i\Omega\,\rho_{RL} & 0
\end{pmatrix}.
\]

\paragraph{Jump $L_+=\sqrt{k_{LR}}\sigma_+$.}
Compute
\[
L_+\rho L_+^\dagger=k_{LR}\sigma_+\rho\sigma_-=
k_{LR}\begin{pmatrix}0&0\\0&\rho_{LL}\end{pmatrix},
\qquad
L_+^\dagger L_+=k_{LR}\sigma_-\sigma_+=
k_{LR}\begin{pmatrix}1&0\\0&0\end{pmatrix}.
\]
Thus
\[
\mathcal{D}[L_+]\rho
=
L_+\rho L_+^\dagger-\frac12\{L_+^\dagger L_+,\rho\}
=
\begin{pmatrix}
-k_{LR}\rho_{LL} & -\frac{k_{LR}}{2}\rho_{LR}\\
-\frac{k_{LR}}{2}\rho_{RL} & +k_{LR}\rho_{LL}
\end{pmatrix}.
\]

\paragraph{Jump $L_-=\sqrt{k_{RL}}\sigma_-$.}
Similarly,
\[
\mathcal{D}[L_-]\rho=
\begin{pmatrix}
+k_{RL}\rho_{RR} & -\frac{k_{RL}}{2}\rho_{LR}\\
-\frac{k_{RL}}{2}\rho_{RL} & -k_{RL}\rho_{RR}
\end{pmatrix}.
\]

\paragraph{Dephasing $L_d=\sqrt{\kappa}\sigma_z$.}
Since $\sigma_z^2=\id$,
\[
\mathcal{D}[L_d]\rho=
\kappa\sigma_z\rho\sigma_z-\kappa\rho=
\begin{pmatrix}
0 & -2\kappa\rho_{LR}\\
-2\kappa\rho_{RL} & 0
\end{pmatrix}.
\]

\paragraph{Combine terms.}
Summing Hamiltonian and dissipators gives
\[
\dot\rho_{LL}=-k_{LR}\rho_{LL}+k_{RL}\rho_{RR},\qquad
\dot\rho_{RR}=+k_{LR}\rho_{LL}-k_{RL}\rho_{RR},
\]
and
\[
\dot\rho_{LR}=i\Omega\rho_{LR}-\frac{k_{LR}+k_{RL}}{2}\rho_{LR}-2\kappa\rho_{LR}
=\left(i\Omega-\frac{\Gamma}{2}-2\kappa\right)\rho_{LR},
\]
which is Theorem~\ref{thm:evolution}.

\section{Proof details for the Kraus equivalence}\label{app:kraus_proof}

\paragraph{Notation.} We write $\rho$ for $\rho_{\mathrm{mod}}$ throughout.

This appendix provides the explicit matrix-element computation for Theorem~\ref{thm:kraus_equivalence}.

Write $\rho(0)=\begin{pmatrix}a&c\\c^*&b\end{pmatrix}$ with $a+b=1$.
Let $\mathcal{E}_t(\rho)=\sum_{j=0}^3 E_j\rho E_j^\dagger$.
Using \eqref{eq:E01}--\eqref{eq:E23} one finds
\begin{align*}
E_0\rho E_0^\dagger &= p_L^\infty\begin{pmatrix} a & c\sqrt{1-\lambda}\\ c^*\sqrt{1-\lambda} & b(1-\lambda)\end{pmatrix},\\
E_1\rho E_1^\dagger &= p_L^\infty\begin{pmatrix} \lambda b & 0\\ 0 & 0\end{pmatrix},\\
E_2\rho E_2^\dagger &= p_R^\infty\begin{pmatrix} (1-\lambda)a & c\sqrt{1-\lambda}\\ c^*\sqrt{1-\lambda} & b\end{pmatrix},\\
E_3\rho E_3^\dagger &= p_R^\infty\begin{pmatrix} 0 & 0\\ 0 & \lambda a\end{pmatrix}.
\end{align*}
Summing yields
\[
(\mathcal{E}_t(\rho))_{LL}=(1-\lambda)a+\lambda p_L^\infty,\qquad
(\mathcal{E}_t(\rho))_{LR}=\sqrt{1-\lambda}\,c.
\]

Next apply $\mathcal{Z}_t(\rho)=\sum_{a=0}^1 F_a\rho F_a^\dagger$ with \eqref{eq:Fa}.
Since $F_0\propto \id$ and $F_1\propto \sigma_z$,
\[
\mathcal{Z}_t(\rho)=\frac{1+\eta}{2}\rho+\frac{1-\eta}{2}\sigma_z\rho\sigma_z,
\]
so diagonal entries are unchanged and $(\mathcal{Z}_t(\rho))_{LR}=\eta\rho_{LR}$.
Finally, $\mathcal{U}_t(\rho)=U\rho U^\dagger$ preserves diagonal entries and multiplies $\rho_{LR}$ by $e^{i\Omega t}$.
Combining $\Lambda_t=\mathcal{U}_t\circ \mathcal{Z}_t\circ \mathcal{E}_t$ gives
\[
\rho_{LL}(t)=(1-\lambda(t))\,\rho_{LL}(0)+\lambda(t)p_L^\infty
=p_L^\infty+(\rho_{LL}(0)-p_L^\infty)e^{-\Gamma t},
\]
since $1-\lambda(t)=e^{-\Gamma t}$, and
\[
\rho_{LR}(t)=e^{i\Omega t}\eta(t)\sqrt{1-\lambda(t)}\,\rho_{LR}(0)
=\rho_{LR}(0)\exp\!\left[\left(i\Omega-\frac{\Gamma}{2}-2\kappa\right)t\right],
\]
since $\eta(t)=e^{-2\kappa t}$ and $\sqrt{1-\lambda(t)}=e^{-\Gamma t/2}$.
This reproduces \eqref{eq:kraus_pop}--\eqref{eq:kraus_coh}.

\section{Why pure Hamiltonian evolution cannot model switching}\label{app:hamiltonian_fails}

This appendix formalizes why a purely Hamiltonian evolution does not change diagonal populations when $H$ is diagonal in the $\{\ket{L},\ket{R}\}$ basis.

\begin{theorem}[Diagonal Hamiltonians preserve diagonal populations]\label{thm:H_fails}
Let $H=c_0\id+c_z\sigma_z$ and let $\rho=\mathrm{diag}(p_L,p_R)$.
Then the von Neumann equation $\dot\rho=-(i/\hbar)[H,\rho]$ gives $\dot\rho=0$.
\end{theorem}

\begin{proof}
Both $H$ and $\rho$ are diagonal matrices, hence they commute: $[H,\rho]=0$.
\end{proof}

\begin{remark}
Geometrically, $H\propto\sigma_z$ generates rotations about the Bloch $z$-axis.
Diagonal states lie on that axis and are invariant under such rotations.
Population transfer requires either off-diagonal Hamiltonian terms ($\sigma_x,\sigma_y$ components) or dissipative jump terms (GKSL).
\end{remark}

\subsection{Non-Hermitian ``dissipative Hamiltonians'' lead to nonlinear dynamics}

A common heuristic is to use $H_{\mathrm{diss}}=H-iG$ with $G\succeq0$, and then renormalize:
\[
\rho(t)=\frac{e^{-iH_{\mathrm{diss}}t/\hbar}\rho(0)e^{+iH_{\mathrm{diss}}^\dagger t/\hbar}}{\Tr(\cdots)}.
\]
This yields a nonlinear evolution.
\begin{theorem}[Renormalized non-Hermitian evolution is nonlinear]\label{thm:nonherm}
The above normalized evolution satisfies
\begin{equation}\label{eq:nonherm_nonlinear}
\frac{\dd\rho}{\dd t}
=-\frac{i}{\hbar}[H,\rho]-\frac{1}{\hbar}\{G,\rho\}
+\frac{2}{\hbar}\Tr(G\rho)\,\rho,
\end{equation}
whose last term is quadratic in $\rho$.
\end{theorem}

\begin{proof}
Differentiate the unnormalized $\tilde\rho(t)=e^{-iH_{\mathrm{diss}}t/\hbar}\rho(0)e^{+iH_{\mathrm{diss}}^\dagger t/\hbar}$ and apply the quotient rule to $\rho=\tilde\rho/\Tr(\tilde\rho)$.
\end{proof}

\begin{remark}
Such normalized non-Hermitian dynamics is not a linear CPTP map on density matrices; it corresponds to a conditional (post-selected) ``no-jump'' evolution in quantum-trajectory language, not an unconditional dissipative dynamics.
\end{remark}

\section{Symmetry and the energy splitting parameter}\label{app:symmetry}

In symmetric Duffing parameters (no bias; symmetric potential), one typically expects equal long-time occupations and no meaningful energy splitting between wells.
We record a simple symmetry argument.

For the symmetric Duffing oscillator with potential $V(x)=-\alpha x^2/2+\beta x^4/4$ (which satisfies $V(-x)=V(x)$ for any $\alpha,\beta>0$), the wells centered at $x=\pm\sqrt{\alpha/\beta}$ have equal depths, implying $\Delta E=0$.
This yields $\Omega=0$ throughout, ensuring the Bloch half-disk geometry remains closed under GKSL evolution.
Asymmetric forcing or potential bias would break this symmetry and introduce nonzero $\Omega$.

\begin{proposition}[Parity symmetry of the standard Duffing potential]
For $V(x)=-(\alpha/2)x^2+(\beta/4)x^4$, one has $V(-x)=V(x)$.
\end{proposition}

\begin{definition}[Conditional energy averages]
Define
\[
\expect{E}_L=\frac{1}{|\Omega_L|}\int_{\Omega_L}E(t)\,\dd t,\quad \Omega_L=\{t:\ x(t)<0\},
\qquad
\expect{E}_R=\frac{1}{|\Omega_R|}\int_{\Omega_R}E(t)\,\dd t,\quad \Omega_R=\{t:\ x(t)>0\},
\]
where $E(t)=\frac12\dot x(t)^2+V(x(t))$.
\end{definition}

\begin{theorem}[Symmetry implies zero conditional energy splitting]\label{thm:DeltaE_zero}
If the stationary statistics of $(x,\dot x)$ are symmetric under $(x,\dot x)\mapsto(-x,-\dot x)$, then $\expect{E}_L=\expect{E}_R$ and thus any conditional energy splitting vanishes.
\end{theorem}

\begin{proof}
Since $E(-x,-\dot x)=E(x,\dot x)$ and the parity map exchanges the events $\{x<0\}$ and $\{x>0\}$ under the symmetry assumption, the conditional distributions are mapped into each other and the conditional expectations coincide.
\end{proof}

\begin{remark}[Finite-time ergodicity and experimental validation]
Theorem~\ref{thm:DeltaE_zero} assumes that the stationary statistics are fully symmetric and that the system has reached ergodic equilibrium.
In practice, finite-time numerical simulations may not fully explore the attractor, particularly in chaotic systems with complex basin structures or slow mixing.
Consequently, empirical estimates of $\expect{E}_L$ and $\expect{E}_R$ may exhibit small asymmetries that are artifacts of incomplete sampling rather than genuine physical asymmetry.
When experimental data suggest nonzero energy splitting or asymmetric dwell-time statistics (see Section~\ref{sec:markov_tests}), it is prudent to verify convergence by extending the simulation duration and checking stability of statistical estimates, test multiple initial conditions to assess sensitivity to transient dynamics, and consider higher-resolution diagnostics (e.g., phase-space density plots on the Poincar\'e section) to confirm whether the observed asymmetry is systematic or sampling-related.
These checks help distinguish true symmetry breaking from finite-time numerical artifacts.
\end{remark}

\section{Summary of key formulas}\label{app:formulas}

For convenience we collect a few formulas used throughout the paper.
We distinguish $\rho_{\mathrm{emb}}$ (embedded state from phase-space data) and $\rho_{\mathrm{mod}}$ (model state governed by GKSL).

\paragraph{Embedded state (from phase-space data).}
\begin{center}
\renewcommand{\arraystretch}{1.4}
\begin{tabular}{@{}ll@{}}
\toprule
Quantity & Formula\\
\midrule
Definition & $\rho_{\mathrm{emb}}(t;\varepsilon,\Delta)=\int \mu_t(z)\,|\psi(z)\rangle\langle\psi(z)|\,dz$\\
Off-diagonal element & $[\rho_{\mathrm{emb}}]_{LR} = c(t;\varepsilon) = \langle\sqrt{w_L w_R}\rangle$\\
Half-disk constraint & $c^2\le p_L(1-p_L)$, with $m_x=2c\ge0$ and $m_x^2+m_z^2\le1$\\
\bottomrule
\end{tabular}
\end{center}

\paragraph{Model state (GKSL dynamics).}
\begin{center}
\renewcommand{\arraystretch}{1.4}
\begin{tabular}{@{}ll@{}}
\toprule
Quantity & Formula\\
\midrule
GKSL equation & $\dot\rho_{\mathrm{mod}}=-(i/\hbar)[H,\rho_{\mathrm{mod}}]+\sum_k(L_k\rho_{\mathrm{mod}} L_k^\dagger-\frac12\{L_k^\dagger L_k,\rho_{\mathrm{mod}}\})$\\
Jump operators & $L_+=\sqrt{k_{LR}}\sigma_+$, $L_-=\sqrt{k_{RL}}\sigma_-$\\
Relaxation rate & $\Gamma=k_{LR}+k_{RL}$\\
Steady state & $p_L^\infty=k_{RL}/\Gamma$, $p_R^\infty=k_{LR}/\Gamma$; $[\rho_{\mathrm{mod}}^\infty]_{LR}=0$\\
Population solution & $p_L(t)=p_L^\infty+(p_0-p_L^\infty)e^{-\Gamma t}$\\
Off-diagonal solution & $[\rho_{\mathrm{mod}}]_{LR}(t)=[\rho_{\mathrm{mod}}]_{LR}(0)e^{(i\Omega-\Gamma/2-2\kappa)t}$\\
Discrete-time $P_{ij}$ & $P_{LR}=p_R^\infty(1-e^{-\Gamma T})$, $P_{RL}=p_L^\infty(1-e^{-\Gamma T})$\\
Rate inversion & $\hat\Gamma=-(1/T)\ln(1-\hat P_{LR}-\hat P_{RL})$, then \eqref{eq:k_hat}\\
\bottomrule
\end{tabular}
\end{center}

\section{Notation and terminology}\label{app:notation}

This appendix provides a comprehensive reference for all key symbols, operators, and terminology used throughout the paper, organized by category. For first appearance, we refer to the relevant equation, definition, theorem, or section.

\subsection{Physical system and observables}

\begin{center}
\renewcommand{\arraystretch}{1.5}
\begin{tabular}{@{}p{3cm}p{7cm}p{3cm}@{}}
\toprule
\textbf{Symbol} & \textbf{Description} & \textbf{First appearance}\\
\midrule
$x$ & Displacement of Duffing oscillator & Eq.~\eqref{eq:duffing}\\
$\dot{x}$ & Velocity of Duffing oscillator & Eq.~\eqref{eq:duffing}\\
$\delta$ & Damping coefficient ($\delta>0$) & Eq.~\eqref{eq:duffing}\\
$\alpha$ & Linear restoring force coefficient ($\alpha>0$) & Eq.~\eqref{eq:duffing}\\
$\beta$ & Cubic nonlinearity coefficient ($\beta>0$) & Eq.~\eqref{eq:duffing}\\
$\gamma_0$ & Forcing amplitude & Eq.~\eqref{eq:duffing}\\
$\omega$ & Forcing frequency & Eq.~\eqref{eq:duffing}\\
$V(x)$ & Potential energy function & Eq.~\eqref{eq:potential}\\
$x_\pm$ & Stable equilibria at $\pm\sqrt{\alpha/\beta}$ & Eq.~\eqref{eq:potential}\\
$\Delta V$ & Barrier height $\alpha^2/(4\beta)$ & Eq.~\eqref{eq:potential}\\
$D$ & Noise intensity (stochastic extension) & Eq.~\eqref{eq:duffing_noise}\\
$T$ & Forcing period $2\pi/\omega$ & Sec.~\ref{sec:methods}\\
$t_n$ & Sampling times $nT$ (Poincar\'e section) & Sec.~\ref{sec:methods}\\
$z$ & Phase-space point $(x,\dot{x})$ on Poincar\'e section & Def.~\ref{def:soft_partition}\\
$E(t)$ & Total energy $\frac12\dot{x}^2+V(x)$ & App.~\ref{app:symmetry}\\
$\Omega_L$ & Time domain when trajectory is in left well, $\{t: x(t)<0\}$ & App.~\ref{app:symmetry}\\
$\Omega_R$ & Time domain when trajectory is in right well, $\{t: x(t)>0\}$ & App.~\ref{app:symmetry}\\
\bottomrule
\end{tabular}
\end{center}

\subsection{Symbol sequences and coarse-graining}

\begin{center}
\renewcommand{\arraystretch}{1.5}
\begin{tabular}{@{}p{3cm}p{7cm}p{3cm}@{}}
\toprule
\textbf{Symbol} & \textbf{Description} & \textbf{First appearance}\\
\midrule
$S_n$ & Binary symbol at time $t_n$ ($S_n\in\{L,R\}$) & Def.~\ref{def:symbol}\\
$\{S_n\}$ & Discrete-time symbol sequence & Def.~\ref{def:symbol}\\
$L$ & Left-well symbol (typically $x<0$) & Def.~\ref{def:symbol}\\
$R$ & Right-well symbol (typically $x>0$) & Def.~\ref{def:symbol}\\
$g(z)$ & Signed boundary function (separatrix approximation) & Def.~\ref{def:soft_partition}\\
$\varepsilon$ & Fuzziness parameter for soft partition ($\varepsilon>0$) & Def.~\ref{def:soft_partition}\\
$w_L(z)$ & Soft membership function for left well & Eq.~\eqref{eq:soft_wL}\\
$w_R(z)$ & Soft membership function for right well ($w_R=1-w_L$) & Eq.~\eqref{eq:soft_wL}\\
$\mu_t(z)$ & Phase-space probability density or empirical measure & Def.~\ref{def:rho_from_mu}\\
$\Delta$ & Window length for empirical measure & Def.~\ref{def:window_measure}\\
$N$ & Number of Poincar\'e samples & Sec.~\ref{sec:numerics_pipeline}\\
\bottomrule
\end{tabular}
\end{center}

\subsection{Quantum-inspired two-state representation}

\begin{center}
\renewcommand{\arraystretch}{1.5}
\begin{tabular}{@{}p{3cm}p{7cm}p{3cm}@{}}
\toprule
\textbf{Symbol} & \textbf{Description} & \textbf{First appearance}\\
\midrule
$\ket{L}$ & Computational basis state for left well & Sec.~\ref{sec:coarse_graining}\\
$\ket{R}$ & Computational basis state for right well & Sec.~\ref{sec:coarse_graining}\\
$\ket{\psi(z)}$ & Pointwise embedded two-state vector & Eq.~\eqref{eq:psi_z}\\
$\rho$ & Generic $2\times2$ density matrix & Eq.~\eqref{eq:bloch}\\
$\rho_{\mathrm{emb}}$ & Embedded state from phase-space data & Eq.~\eqref{eq:rho_from_mu}\\
$\rho_{\mathrm{mod}}$ & Model state governed by GKSL dynamics & Sec.~\ref{sec:gksl_model}\\
$\rho_{LL}$ & Diagonal element (left-well population) & Thm.~\ref{thm:evolution}\\
$\rho_{RR}$ & Diagonal element (right-well population) & Thm.~\ref{thm:evolution}\\
$\rho_{LR}$ & Off-diagonal element (coherence) & Thm.~\ref{thm:evolution}\\
$\rho_{RL}$ & Off-diagonal element ($\rho_{RL}=\rho_{LR}^*$) & Thm.~\ref{thm:evolution}\\
$\rho^\infty$ & Steady-state density matrix & Thm.~\ref{thm:steady}\\
$p_L(t)$ & Population in left well at time $t$ & Prop.~\ref{prop:rho_elements_overlap}\\
$p_R(t)$ & Population in right well at time $t$ ($p_R=1-p_L$) & Prop.~\ref{prop:rho_elements_overlap}\\
$p_L^\infty$ & Steady-state left-well population $k_{RL}/\Gamma$ & Thm.~\ref{thm:steady}\\
$p_R^\infty$ & Steady-state right-well population $k_{LR}/\Gamma$ & Thm.~\ref{thm:steady}\\
$c(t)$ & Overlap parameter $\langle\sqrt{w_L w_R}\rangle$ & Prop.~\ref{prop:rho_elements_overlap}\\
$c(\varepsilon)$ & Overlap parameter as function of fuzziness & Eq.~\eqref{eq:c_hat}\\
$\hat{c}(\varepsilon)$ & Empirical estimate of overlap parameter & Eq.~\eqref{eq:c_hat}\\
\bottomrule
\end{tabular}
\end{center}

\subsection{Pauli matrices and Bloch representation}

\begin{center}
\renewcommand{\arraystretch}{1.5}
\begin{tabular}{@{}p{3cm}p{7cm}p{3cm}@{}}
\toprule
\textbf{Symbol} & \textbf{Description} & \textbf{First appearance}\\
\midrule
$\sigma_x$ & Pauli-X matrix (bit-flip) & Eq.~\eqref{eq:pauli}\\
$\sigma_y$ & Pauli-Y matrix & Eq.~\eqref{eq:pauli}\\
$\sigma_z$ & Pauli-Z matrix (population difference) & Eq.~\eqref{eq:pauli}\\
$\sigma_+$ & Raising operator $\ketbra{R}{L}$ & Eq.~\eqref{eq:sigma_pm}\\
$\sigma_-$ & Lowering operator $\ketbra{L}{R}$ & Eq.~\eqref{eq:sigma_pm}\\
$\bm{m}$ & Bloch vector $(m_x,m_y,m_z)$ & Eq.~\eqref{eq:bloch}\\
$m_x$ & Bloch $x$-component ($m_x=2c$ for real states) & Eq.~\eqref{eq:bloch}\\
$m_y$ & Bloch $y$-component ($m_y=0$ for symmetric case) & Eq.~\eqref{eq:bloch}\\
$m_z$ & Bloch $z$-component ($m_z=p_L-p_R=2p_L-1$) & Eq.~\eqref{eq:bloch}\\
$m_z^\infty$ & Steady-state $z$-component $(k_{RL}-k_{LR})/\Gamma$ & Eq.~\eqref{eq:mz}\\
$\id$ & $2\times2$ identity matrix & Eq.~\eqref{eq:bloch}\\
\bottomrule
\end{tabular}
\end{center}

\subsection{GKSL generator and rates}

\begin{center}
\renewcommand{\arraystretch}{1.5}
\begin{tabular}{@{}p{3cm}p{7cm}p{3cm}@{}}
\toprule
\textbf{Symbol} & \textbf{Description} & \textbf{First appearance}\\
\midrule
$H$ & Hamiltonian operator $-(\Delta E/2)\sigma_z$ & Eq.~\eqref{eq:H_model}\\
$\Delta E$ & Energy splitting between wells & Eq.~\eqref{eq:H_model}\\
$\Omega$ & Rabi-like frequency $\Delta E/\hbar$ & Def.~\ref{def:model}\\
$\hbar$ & Reduced Planck constant (formal parameter) & Eq.~\eqref{eq:lindblad_general}\\
$L_k$ & Generic Lindblad jump operator & Eq.~\eqref{eq:lindblad_general}\\
$L_+$ & Jump operator $\sqrt{k_{LR}}\,\sigma_+$ (L$\to$R) & Eq.~\eqref{eq:L_plus}\\
$L_-$ & Jump operator $\sqrt{k_{RL}}\,\sigma_-$ (R$\to$L) & Eq.~\eqref{eq:L_minus}\\
$L_d$ & Dephasing operator $\sqrt{\kappa}\,\sigma_z$ & Eq.~\eqref{eq:L_d}\\
$k_{LR}$ & Switching rate from left to right well & Def.~\ref{def:model}\\
$k_{RL}$ & Switching rate from right to left well & Def.~\ref{def:model}\\
$\Gamma$ & Total relaxation rate $k_{LR}+k_{RL}$ & Thm.~\ref{thm:evolution}\\
$\kappa$ & Pure dephasing rate & Eq.~\eqref{eq:L_d}\\
$\tau_{\mathrm{rel}}$ & Population relaxation time $1/\Gamma$ & Sec.~\ref{sec:solutions}\\
$\tau_{\mathrm{off}}$ & Off-diagonal decay time $1/(\Gamma/2+2\kappa)$ & Sec.~\ref{sec:solutions}\\
$\hat{\Gamma}$ & Estimated total relaxation rate & Eq.~\eqref{eq:Gamma_hat}\\
$\hat{k}_{LR}$ & Estimated L$\to$R switching rate & Eq.~\eqref{eq:k_hat}\\
$\hat{k}_{RL}$ & Estimated R$\to$L switching rate & Eq.~\eqref{eq:k_hat}\\
\bottomrule
\end{tabular}
\end{center}

\subsection{Discrete-time transition probabilities}

\begin{center}
\renewcommand{\arraystretch}{1.5}
\begin{tabular}{@{}p{3cm}p{7cm}p{3cm}@{}}
\toprule
\textbf{Symbol} & \textbf{Description} & \textbf{First appearance}\\
\midrule
$\mathbf{P}$ & One-step transition probability matrix & Eq.~\eqref{eq:transition_matrix}\\
$P_{ij}$ & Transition probability from state $i$ to state $j$ & Eq.~\eqref{eq:Pij}\\
$P_{LR}$ & One-step transition probability L$\to$R & Eq.~\eqref{eq:Pij}\\
$P_{RL}$ & One-step transition probability R$\to$L & Eq.~\eqref{eq:Pij}\\
$\hat{P}_{ij}$ & Empirical estimate of $P_{ij}$ & Sec.~\ref{sec:markov_tests}\\
$\widehat{\mathbf{P}}$ & Empirical one-step transition matrix & Sec.~\ref{sec:markov_tests}\\
$\widehat{\mathbf{P}}^{(2)}$ & Empirical two-step transition matrix & Sec.~\ref{sec:markov_tests}\\
$N_{ij}$ & Transition count from state $i$ to state $j$ & Sec.~\ref{sec:markov_tests}\\
$N_{kij}$ & Second-order transition count $k\to i\to j$ & Sec.~\ref{sec:markov_tests}\\
$p_L^{(n)}$ & Left-well population at discrete time $n$ & Eq.~\eqref{eq:discrete_update}\\
\bottomrule
\end{tabular}
\end{center}

\subsection{Kraus operators}

\begin{center}
\renewcommand{\arraystretch}{1.5}
\begin{tabular}{@{}p{3cm}p{7cm}p{3cm}@{}}
\toprule
\textbf{Symbol} & \textbf{Description} & \textbf{First appearance}\\
\midrule
$K_\ell(t)$ & Generic Kraus operator & Sec.~\ref{sec:kraus}\\
$\Lambda_t$ & CPTP map $\rho(0)\mapsto\rho(t)$ & Sec.~\ref{sec:kraus}\\
$\lambda(t)$ & Relaxation parameter $1-e^{-\Gamma t}$ & Sec.~\ref{sec:kraus}\\
$\eta(t)$ & Dephasing parameter $e^{-2\kappa t}$ & Sec.~\ref{sec:kraus}\\
$U(t)$ & Unitary evolution operator $e^{-iHt/\hbar}$ & Sec.~\ref{sec:kraus}\\
$E_j(t)$ & GAD (generalized amplitude damping) Kraus operators & Eq.~\eqref{eq:E01}, \eqref{eq:E23}\\
$F_a(t)$ & Phase-damping Kraus operators & Eq.~\eqref{eq:Fa}\\
$M_{aj}(t)$ & Composite Kraus operators $U(t)F_a(t)E_j(t)$ & Eq.~\eqref{eq:M_aj_def}\\
\bottomrule
\end{tabular}
\end{center}

\subsection{Diagnostic tests and statistics}

\begin{center}
\renewcommand{\arraystretch}{1.5}
\begin{tabular}{@{}p{3cm}p{7cm}p{3cm}@{}}
\toprule
\textbf{Symbol} & \textbf{Description} & \textbf{First appearance}\\
\midrule
$G$ & G-test statistic (likelihood ratio) & Eq.~\eqref{eq:g_test}\\
$d$ & Dimension of state space (number of states) & Eq.~\eqref{eq:order_test_df}\\
$\widehat{P}_{ij\mid k}$ & Conditional transition probability given previous state $k$ & Sec.~\ref{sec:markov_tests}\\
$\Delta_{\mathrm{CK}}$ & Chapman-Kolmogorov discrepancy $\|\widehat{\mathbf{P}}^{(2)}-\widehat{\mathbf{P}}^2\|_F$ & Sec.~\ref{sec:markov_tests}\\
$p_{\mathrm{CK}}$ & Chapman-Kolmogorov test $p$-value & Table~\ref{tab:bootstrap_ck_test}\\
$R_L$ & Run length (dwell time) in left well & Sec.~\ref{sec:markov_tests}\\
$R_R$ & Run length (dwell time) in right well & Sec.~\ref{sec:markov_tests}\\
$W$ & Number of windows for stationarity test & Sec.~\ref{sec:markov_tests}\\
$\hat{P}^{(w)}_{ij}$ & Transition probability in window $w$ & Sec.~\ref{sec:markov_tests}\\
$B$ & Number of bootstrap replicates & Sec.~\ref{sec:block_bootstrap}\\
$m$ & Number of blocks in bootstrap resampling & Table~\ref{tab:block_bootstrap}\\
$\ell$ & Block length for block bootstrap & Eq.~\eqref{eq:block_length}\\
$\tau_{\mathrm{int}}$ & Integrated autocorrelation time & Sec.~\ref{sec:block_bootstrap}\\
$\mathrm{df}$ & Degrees of freedom for $\chi^2$ test & Eq.~\eqref{eq:order_test_df}\\
\bottomrule
\end{tabular}
\end{center}

\subsection{Acronyms and abbreviations}

\begin{center}
\renewcommand{\arraystretch}{1.5}
\begin{tabular}{@{}p{3cm}p{10cm}@{}}
\toprule
\textbf{Acronym} & \textbf{Meaning}\\
\midrule
GKSL & Gorini-Kossakowski-Sudarshan-Lindblad (master equation)\\
CPTP & Completely Positive and Trace-Preserving (quantum channel)\\
GAD & Generalized Amplitude Damping (quantum channel)\\
ODE & Ordinary Differential Equation\\
SDE & Stochastic Differential Equation\\
\bottomrule
\end{tabular}
\end{center}

\subsection{Mathematical operators and notation}

\begin{center}
\renewcommand{\arraystretch}{1.5}
\begin{tabular}{@{}p{3cm}p{7cm}p{3cm}@{}}
\toprule
\textbf{Symbol} & \textbf{Description} & \textbf{First appearance}\\
\midrule
$\Tr$ & Trace operator & Sec.~2 (custom command)\\
$\dd$ & Differential operator (upright d) & Sec.~2 (custom command)\\
$\ket{\cdot}$ & Ket (column) vector in Dirac notation & Sec.~2 (custom command)\\
$\bra{\cdot}$ & Bra (row) vector in Dirac notation & Sec.~2 (custom command)\\
$\ketbra{\cdot}{\cdot}$ & Outer product $\ket{\cdot}\bra{\cdot}$ & Sec.~2 (custom command)\\
$\expect{\cdot}$ & Expectation value & Sec.~2 (custom command)\\
$\norm{\cdot}$ & Norm (default: Euclidean or Frobenius) & Sec.~2 (custom command)\\
$\norm{\cdot}_F$ & Frobenius norm & Sec.~\ref{sec:markov_tests}\\
$[A,B]$ & Commutator $AB-BA$ & Eq.~\eqref{eq:lindblad_general}\\
$\{A,B\}$ & Anticommutator $AB+BA$ & Eq.~\eqref{eq:lindblad_general}\\
$A^\dagger$ & Hermitian adjoint (conjugate transpose) & Eq.~\eqref{eq:lindblad_general}\\
$A^*$ & Complex conjugate (Hermitian adjoint for scalars) & Throughout\\
$A^\infty$ & Steady-state (long-time) value & Throughout\\
\bottomrule
\end{tabular}
\end{center}

\end{document}